\newif\ifreport\reporttrue

\newif\ifnotreport
\ifreport
\notreportfalse
\else
\notreporttrue
\fi

\ifreport
\documentclass[sigconf]{acmart}
\setcopyright{none}
\acmConference[ASE '22]{37th IEEE/ACM International Conference on Automated Software Engineering}{October 10--14, 2022}{Rochester, MI, USA}
\renewcommand\footnotetextcopyrightpermission[1]{}
\else
\documentclass[sigconf]{acmart}
\copyrightyear{2022} 
\acmYear{2022} 
\setcopyright{rightsretained} 
\acmConference[ASE '22]{37th IEEE/ACM International Conference on Automated Software Engineering}{October 10--14, 2022}{Rochester, MI, USA}
\acmBooktitle{37th IEEE/ACM International Conference on Automated Software Engineering (ASE '22), October 10--14, 2022, Rochester, MI, USA}
\acmDOI{10.1145/3551349.3561152}
\acmISBN{978-1-4503-9475-8/22/10}
\fi

\usepackage{booktabs}   
\usepackage{subcaption} 

\usepackage{colortbl}
\usepackage[utf8]{inputenc}
\usepackage{color}
\usepackage[linesnumbered,ruled,vlined]{algorithm2e}
\usepackage{tikz}
\usetikzlibrary{arrows.meta}
\usetikzlibrary{patterns}
\usepackage{acronym}
\usepackage{pifont}
\usepackage[override]{cmtt}
\usepackage{ifplatform}
\usepackage{multirow}
\usepackage{url}
\usepackage{pgfplots, verbatim}
\usepackage{multicol}
\usepackage[htt]{hyphenat}
\usepackage{tabularx}
\usepackage{tablefootnote}
\usepackage{threeparttable}
\usepackage{listings}
\lstdefinelanguage{ASM}{
    morekeywords= [1]{
        r0, r1, r2, r3, r4, r5, r6, r7, r8, r9, r10, r11, r12,
        sp, lr},
    morekeywords = [3]{b, ble, blt, bne,ldr, str, cmp, moveq, sub, add, pop, push, mov, bl, bic, pc},
    sensitive=false, 
    morecomment=[l]{//}, 
    morecomment=[s]{/*}{*/}, 
    morestring=[b]", 
    keywordstyle = [1]\color{red},
    keywordstyle = [3]\color{blue}
}

\newcommand{\tool}{{\textsc{Cornucopia}}\xspace}
\newcommand{\ourtool}{\tool}
\newcommand{\angr}{{\sc angr}\xspace}
\newcommand{\qemu}{{\sc qemu}\xspace}
\newcommand{\radare}{{\sc radare}\xspace}
\newcommand{\ghidra}{{\sc Ghidra}\xspace}
\newcommand{\idapro}{{\sc ida}\xspace}
\newcommand{\tigress}{{\sc tigress}\xspace}
\newcommand{\AFL}{{\sc AFL++}\xspace}
\newcommand{\debin}{{\sc Debin}\xspace}
\newcommand{\stateformer}{{\sc StateFormer}\xspace}

\newcommand{\obfusllvm}{{\sc ob-llvm}\xspace}
\newcommand{\bintuner}{{\textsc{BinTuner}}\xspace}
\newcommand{\bindiff}{$BD_{I}$\xspace}
\newcommand{\ncdo}{$N_{o}$\xspace}
\newcommand{\ncdavg}{$N_{a}$\xspace}
\newcommand{\ncdmin}{$N_{m}$\xspace}
\newcommand{\pwavg}{$P_{a}$\xspace}
\newcommand{\pwmin}{$P_{m}$\xspace}
\newcommand{\funchash}{$F_{h}$\xspace}
\newcommand{\proxyp}{$Proxy_{p}$\xspace}
\newcommand{\failures}{$F_{s}$\xspace}
\newcommand{\failurem}{$F_{m}$\xspace}
\newcommand{\successa}{$S_{a}$\xspace}
\newcommand{\fitserver}{$Fit_{Ser}$\xspace}
\newcommand{\cosinesim}{$CS_{I}$\xspace}
\newcommand{\xeightsix}{{\sc x86}\xspace}
\newcommand{\xsixfour}{{\sc x64}\xspace}
\newcommand{\arm}{{\sc ARM}\xspace}
\newcommand{\mips}{{\sc MIPS}\xspace}

\newcommand{\asmvec}{{\sc Asm2Vec}\xspace}
\newcommand{\safe}{{\sc SAFE}\xspace}

\newcommand{\cornvsbin}{{\sc 8X}\xspace}
\newcommand{\cornwithbin}{$C_{b}$\xspace}

\acrodef{ISA}{Instruction Set Architecture}
\acrodef{ML}{Machine Learning}
\acrodef{NCD}{Normalized Compression Distance}
\acrodef{PIECEWISE}{Fuzzy Hashing}
\acrodef{WLLVM}{Whole Program LLVM}
\acrodef{DScore}{Difference Score}
\acrodef{llc}{LLVM Bitcode Compiler}
\acrodef{RISC}{Reduced Instruction Set Computer}
\acrodef{CISC}{Complex Instruction Set Computer}
\acrodef{PDF}{Probability Density Function}

\newcommand{\eg}{\textit{e.g.,\ }}
\newcommand{\ie}{\textit{i.e.,\ }}

\newcommand{\code}[1]{
 \textit{\textbf{}}
} 

\newcommand{\tbl}[1]{Table~\ref{#1}}
\newcommand{\sect}[1]{Section~\ref{#1}}
\newcommand{\fig}[1]{Figure~\ref{#1}}
\newcommand{\lst}[1]{Listing~\ref{#1}}

\newcommand{\apdx}[1]{Appendix~\ref{#1}}
\newcommand{\clangversion}{\texttt{clang}\xspace}
\newcommand{\gccversion}{\texttt{gcc}\xspace}

\newcommand{\numllvmpackages}{191\xspace}

\newcommand{\totalclangbinaries}{308,269\xspace}

\newcommand{\avgclangbinnum}{403\xspace}
\newcommand{\crashesangr}{263\xspace}

\newcommand{\medianclangbinnumber}{413\xspace}

\newcommand{\fuzztime}{6\xspace}
\newcommand{\numllvmcrashes}{300\xspace}
\newcommand{\avgcorcbbinaries}{52\xspace}
\newcommand{\avgcorbinaries}{450\xspace}
\newcommand{\avgbintunerbinaries}{48\xspace}

\newcommand{\zenodo}{\url{https://doi.org/10.5281/zenodo.7039858}}

\newcommand{\website}{\url{https://binarygeneration.github.io/}}

\definecolor{akulcolor}{rgb}{0.93, 0.53, 0.18}
\definecolor{vidushcolor}{rgb}{0.0, 0.87, 0.87}

\definecolor{armgreen}{rgb}{0.0, 0.5, 0.0}
\definecolor{darkgreen}{rgb}{0.0, 0.5, 0.0}
\definecolor{darkred}{rgb}{0.5, 0.0, 0.0}
\newcommand{\hlred}[1]{\textcolor{darkred}{\textbf{#1}}}
\newcommand{\hlgreen}[1]{\textcolor{darkgreen}{\textbf{#1}}}

\newcommand{\aravind}[1]{\textcolor{green}{Aravind: #1}}

\newcommand{\akul}[1]{\textcolor{akulcolor}{Akul: #1}}

\def\fullsupp{\ding{51}}
\def\partsupp{\ding{109}}
\def\nosupp{\ding{55}}

\definecolor{thircolor}{rgb}{0.6, 0.4, 0.8}
\definecolor{sixcolor}{rgb}{0.8, 0.58, 0.46}
\definecolor{armcolor}{rgb}{0.55, 0.71, 0.0}
\definecolor{mipscolor}{rgb}{0.0, 0.5, 1.0}

\definecolor{top1}{rgb}{1.0, 0.21, 0.37}
\definecolor{top2}{rgb}{1.0, 0.35, 0.21}
\definecolor{top3}{rgb}{0.89, 0.31, 0.61}

\definecolor{bot3}{rgb}{0.5, 0.0, 0.5}
\definecolor{bot2}{rgb}{0.63, 0.36, 0.94}
\definecolor{bot1}{rgb}{0.59, 0.47, 0.71}

\definecolor{avg}{rgb}{0.28, 0.02, 0.03}
\definecolor{tot}{rgb}{0.0, 0.42, 0.24}

\begin{document}

\setlength{\parskip}{0em}


\ifreport
\title{\ourtool{}: A Framework for Feedback Guided Generation of Binaries}
\subtitle{Extended Report}
\fi
  


\author{Vidush Singhal}
\affiliation{
\institution{Purdue University}
\country{United States}
}
\email{singhav@purdue.edu}

\author{Akul Abhilash Pillai}
\affiliation{
\institution{Purdue University}
\country{United States}
}
\email{pillai23@purdue.edu}

\author{Charitha Saumya}
\affiliation{
\institution{Purdue University}
\country{United States}
}
\email{cgusthin@purdue.edu}

\author{Milind Kulkarni}
\affiliation{
\institution{Purdue University}
\country{United States}
}
\email{milind@purdue.edu}

\author{Aravind Machiry}
\affiliation{
\institution{Purdue University}
\country{United States}
}
\email{amachiry@purdue.edu}

\begin{abstract}
Binary analysis is an important capability required for many security and software engineering applications.
Consequently, there are many binary analysis techniques and tools with varied capabilities.
However, testing these tools requires a large, varied binary dataset with corresponding source-level information.
In this paper, we present~\ourtool{}, an architecture agnostic automated framework that can generate a plethora of binaries from corresponding program source by exploiting compiler optimizations and feedback-guided learning.
Our evaluation shows that~\ourtool{} was able to generate 309K binaries across four architectures (x86, x64, ARM, MIPS) with an average of~\avgclangbinnum{} binaries for each program and outperforms~\bintuner{}~\cite{ren2021unleashing}, a similar technique.
Our experiments revealed issues with the LLVM optimization scheduler resulting in compiler crashes ($\sim$300).
Our evaluation of four popular binary analysis tools \angr,~\ghidra,~\idapro, and~\radare, using~\ourtool generated binaries, revealed various issues with these tools.
Specifically, we found~\crashesangr crashes in~\angr and one memory corruption issue in~\idapro.
Our differential testing on the analysis results revealed various semantic bugs in these tools.
We also tested machine learning tools,~\asmvec,~\safe, and~\debin, that claim to capture binary semantics and show that they perform poorly (\eg \debin F1 score dropped to 12.9\% from reported 63.1\%) on \ourtool generated binaries.
In summary, our exhaustive evaluation shows that~\ourtool{} is an effective mechanism to generate binaries for testing binary analysis techniques effectively.
\end{abstract}

\maketitle
\section{Introduction}
\label{sec:introduction}

Designing proper binary analysis tools is a challenging task.
It requires precisely modeling all the units of the underlying~\ac{ISA}.
Even commonly-used, supposedly-robust tools, such as \qemu,
have bugs in precisely modeling certain instructions~\cite{qemubugs}.
One common approach to designing these tools, especially static analysis tools,
is to perform incremental development~\cite{andriesse2018practical}.
Specifically, instead of painstakingly modeling all the aspects of the underlying ISA,
tool developers model only those instructions and patterns commonly observed in binaries.
These common patterns are highly dependent on which binaries developers consider.
Without evaluating a representative dataset of binaries, some key patterns may be
overlooked, and these tools will thus be less robust.
Similarly,~\ac{ML} techniques~\cite{xue2019machine} used to solve various binary analysis problems
also rely on a varied dataset of binaries for training.
For certain security-critical applications such as malware detection~\cite{singh2021survey},
a misprediction (\ie false negative) by the corresponding~\ac{ML} model can
be disastrous for the security of the underlying system~\cite{kolosnjaji2018adversarial}.
In order to mitigate such issues and to build robust~\ac{ML} models,
it is important to ensure that the training dataset of binaries is sufficiently 
 varied. 

Most existing tools to produce binary datasets~\cite{wang2017ramblr, 280046} use binaries generated using standard optimization
flags (\ie O0, O1, O2, O3, Os, Ofast). 
Unfortunately, the binaries generated using standard optimization flags
frequently miss common idioms~\cite{blackmore2017automatically}.
Consequently, analysis tools developed based on these datasets fail to handle certain idioms, resulting in tool
failures, as evident from a large number of issues in~\angr{}~\cite{shoshitaishvili2016state, angrissues} and~\radare{}~\cite{r2, radrissues}.
These tools enable~\cite{Pang2021SoKAY} important security and software maintenance applications such as Control Flow Integrity (CFI)~\cite{wang2015binary}, Automated Patching~\cite{jeong2017functional}, and Binary Rewriting~\cite{agadakos2019nibbler, wenzl2019hack}.
Failures in these tools impact their usability and delay research progress.
Unfortunately, irrespective of an active open source community, academic researchers
spend considerable time fixing various robustness issues in these tools~\cite{casinghino2019using}.
Similarly,~\ac{ML} tools trained using binaries generated from
standard optimization flags (-Ox)  are shown to perform poorly on binaries
compiled with non-standard optimization flags~\cite{ren2021unleashing}.

We wish to automatically generate well-formed binaries so that binary analysis and~\ac{ML}  tools can use the generated datasets to improve their robustness.
The generated binaries should have associated high-level structures, specifically source code, to enable the creation of ground-truth information (\ie through debug symbols) needed by machine learning tools.
Existing binary-level techniques~\cite{wu2010mimimorphism, szor2001hunting, behera2015different, fang2011multi, 8659363} use semantics-preserving transformations (\eg Register Swapping) to generate several semantically-equivalent binaries from a single binary.
These techniques are primarily designed for program obfuscation and are based on fixed patterns.
Consequently, the number of variants generated for a given binary is limited.
Second, these techniques depend on the ability to perform
static binary rewriting and reassemblable disassembly, which is known to be a hard problem~\cite{wenzl2019hack}.
Third, as mentioned before, we need to have source code or ground truth information corresponding to the generated binaries.
However, generating source code for a given arbitrary binary (\ie decompilation) is known to be a hard problem~\cite{verbeek2020sound}.
Finally, generating semantics preserving transformations requires a
precise model of the underlying~\ac{ISA}, which requires a considerable amount of effort~\cite{armstrong2018detailed, dasgupta2019complete}.
For instance, even a simple register swapping/renaming transformation, such as renaming register~\textbf{RCX} to~\textbf{RDX} in a function, requires knowledge of the ABI.
Specifically, we need to know that the function does not use~\textbf{RCX} or~\textbf{RDX} for its arguments. To determine this, we need to know the number and type (scalar or not) of parameters~\cite{caballero2016type} for the function and the calling convention used by the function. Both are known to be challenging~\cite{grelot2021automation}.

Another class of techniques performs semantics-preserving transformations, but at the source level (\eg\tigress~\cite{tigress}) or IR level (\eg\obfusllvm~\cite{junod2015obfuscator}).
These techniques focus on~\ac{ISA}-agnostic control flow and data
flow related aspects of the program without considering the~\ac{ISA}-dependent 
instruction sequence or patterns used in the resulting binary.
Consequently, these techniques are shown to have~\emph{less} or~\emph{no} impact on 
the generated binary~\cite{madou2006effectiveness}.
\tbl{tbl:relatedcomparision} shows a summary of the existing techniques along with their drawbacks.

In this paper, we focus on the problem of generating {\em large numbers of binaries for a given program}.
We aim to develop a tool that binary analysis framework developers can easily use to test their framework effectively.
Furthermore, We want to have ground truth information (\ie source code and debug information) for all the generated binaries.
We observe that compilers have these precise models of~\ac{ISA} as part of their target code generation component~\cite{srikant2018compiler}.
Most compilers provide various options and target-(in)dependent optimization flags that allow fine-grained control over choices in code generation~\cite{compoptrev}.
Our basic idea is to use these fine-grained optimization flags to generate different binaries.
However, for a given program, not all optimization flags affect the program's binary.
For instance, the flag~\textbf{--x86-use-base-pointer} available in \textbf{clang} does not affect programs
with small local variables.
Although individual flags may be ineffective for certain programs, \textit{combinations} of the flags could generate different binaries~\cite{blackmore2017automatically}.
For a given program, identifying which flag combinations affect the target binary is a combinatorial problem---intractable, especially when there are a large and growing number of flags ($\sim$ 892 usable flags for x86 in~\textbf{clang-12.0}).

In fact, we tried the brute-force approach of enumerating all the combinations of compiler to compile programs of different sizes.
In 12 hours, on average, the brute-force approach was able to generate 197 unique binaries, whereas our approach was able to generate 6,512 (33$\times$) in just 6 hours (half the time).

We present~\tool, an automated, architecture-independent framework for generating a plethora of binaries for a given program.
Given a source package (\eg~\textbf{a2ps.tar.gz}), compiler, and set of all available optimization flags,~\tool{} iteratively learns to produce unique binaries for a given source package by feedback-guided mutation of compiler flags, thus avoiding enumerating all combinations of optimization flags.
A recent work,~\bintuner{}~\cite{ren2021unleashing}, also explores the use of compiler flags to generate different variations of binaries for a given program.
Although it uses a search-based iterative compilation,~\bintuner{}'s goal is not to generate diverse binaries but to generate a binary very different from those generated by general Ox optimization levels.
Furthermore, it requires explicit specification of conflicting compilation options in the form of first-order formulas, which must be specified for every~\ac{ISA} and compiler combination.
This requires an in-depth understanding of various compiler options, which involves considerable effort and conflicts with our requirement for an easy-to-use tool.
Finally, as shown in~\sect{subsub:cornucopavsbintuner}, ~\bintuner{}'s fitness function is inferior to~\tool for generating a plethora of diverse binaries. The latter generated~\cornvsbin{} more binaries than~\bintuner{} in a given time.

{
\scriptsize
\begin{table*}[]
\centering
\begin{threeparttable}
\begin{tabular}{c|c|c|c|c|c}
\toprule
\textbf{Technique}                                                       & \textbf{\begin{tabular}[c]{@{}c@{}}Binaries Have \\ corresponding \\ source code?~\tnote{1}\end{tabular}} & \textbf{\begin{tabular}[c]{@{}c@{}}ISA Specification\\ Not Needed?\end{tabular}} & \textbf{\begin{tabular}[c]{@{}c@{}}Reassemblable\\ Disassembly\\ Not Needed?\end{tabular}} & \textbf{\begin{tabular}[c]{@{}c@{}}Affects \\ Generated\\ Binaries?\end{tabular}} & \textbf{\begin{tabular}[c]{@{}c@{}}Generates \\ Large Number \\ of Binaries?\end{tabular}} \\ \midrule
\rowcolor{black!15} \begin{tabular}[c]{@{}c@{}}Compilation Through Standard \\ Optimization levels~\cite{wang2017ramblr, allstarDataset, 280046}\\ (e.g., O0, O1, O3, etc)\end{tabular} &             \fullsupp                                                                   &      \fullsupp                                                                            &        \fullsupp                                                                                    &     \fullsupp                                                                              &       \nosupp                                                                                     \\ 
\begin{tabular}[c]{@{}c@{}}Source Level\\ Transformations~\cite{tigress, 7781792} \\ (e.g., CFG flattening~\cite{nagra2009surreptitious})\end{tabular}      &       \fullsupp                                                                         &     \fullsupp                                                                             &       \fullsupp                                                                                     &     \partsupp                                                                              &  \partsupp                                                                                          \\ 
\rowcolor{black!15} \begin{tabular}[c]{@{}c@{}}IR level \\ Transformations~\cite{tamboli2014metamorphic, junod2015obfuscator} \\ (e.g., Deadcode Insertion~\cite{barria2016obfuscation})\end{tabular}          &      \nosupp                                                                          &       \fullsupp                                                                           &       \fullsupp                                                                                     &    \partsupp                                                                               &    \partsupp                                                                                        \\ 
\begin{tabular}[c]{@{}c@{}}Binary Level \\ Transformations~\cite{wu2010mimimorphism, szor2001hunting, behera2015different, fang2011multi, 8659363} \\ (e.g., Register renaming~\cite{vdurfina2011generic})\end{tabular}      &   \nosupp                                                                             &     \nosupp                                                                             &     \nosupp                                                                                       &    \fullsupp                                                                               &     \fullsupp                                                                                        \\ 
\rowcolor{black!15} \ourtool{} (Our System)                                                               &   \fullsupp                                                                             &    \fullsupp                                                                              &    \fullsupp                                                                                        &   \fullsupp                                                                                &       \fullsupp                                                                                     \\ \bottomrule
\end{tabular}
\begin{tablenotes}
        \item[1] As mentioned in~\sect{sec:introduction}, this is needed to generate ground truth.
     \end{tablenotes}
\end{threeparttable}
\caption{\emph{Comparison of \ourtool{} to other binary generation techniques}.
	For each of the feature, we indicate whether the technique fully supports (\fullsupp), partially supports (\partsupp), or does not support (\nosupp) it.}
\label{tbl:relatedcomparision}
\end{table*}
}

Our evaluation shows that~\tool{}, in 6 hours, can generate, on average,~\avgclangbinnum binaries per program across all architectures. 
In addition, standard tools for evaluating binary differences show that these binary variants are highly varied
\ifreport
(\ref{fig:bindiffllvmcdf})
\fi
.

Generating a large number of binary variants is only useful if those variants expose interesting behaviors in the software toolchain. The binaries generated by~\tool{} revealed various issues in current static analysis and~\ac{ML} tools, showing the inadequacy of the current methods to make these tools robust.
This shows that~\tool{} generates binaries that can be used to improve the robustness of binary analysis tools.
Additionally, we observed that~\ourtool{} can also be used to test the optimization scheduler in compilers to find issues related to optimization dependencies~\cite{sun2016toward}. We found issues with the LLVM optimization scheduler which resulted in compiler crashes $\sim$300.
In summary, the following are our contributions:
\begin{itemize}
\item We present~\tool{}, a feedback-guided mutation technique to efficiently find sets of compiler optimization flags that produce different binaries for a given application and show that it outperforms~\bintuner (\sect{subsub:cornucopavsbintuner}), a recent approach that tries to find optimization flags resulting in a large binary code difference.
\item Our evaluation shows that~\tool generates a large number of unique binaries for a given program, and these binaries differ significantly from those generated using standard optimization levels (\sect{subsub:bingeneffectiveness}).
\item Our evaluation of existing binary analysis tools and machine learning tools with~\tool generated binaries revealed various robustness issues (\ie \crashesangr crashes in~\angr and one in~\idapro), semantic issues (\sect{subsub:robuststaticanalysis}), and model performance issues (\sect{subsub:robusttest}), demonstrating the utility of~\ourtool{} in testing existing tools.
\item We show that~\ourtool{} is also effective at testing optimization schedulers in compilers by finding issues with the LLVM optimization scheduler which resulted in $\sim$300 compiler crashes .
\item The source code and generated binaries of~\tool{} is available at~\zenodo. Refer to our website for more details:~\website 
\end{itemize}


\section{Overview}
\label{sec:overview}

\begin{figure}[t]
\includegraphics[width=\linewidth]{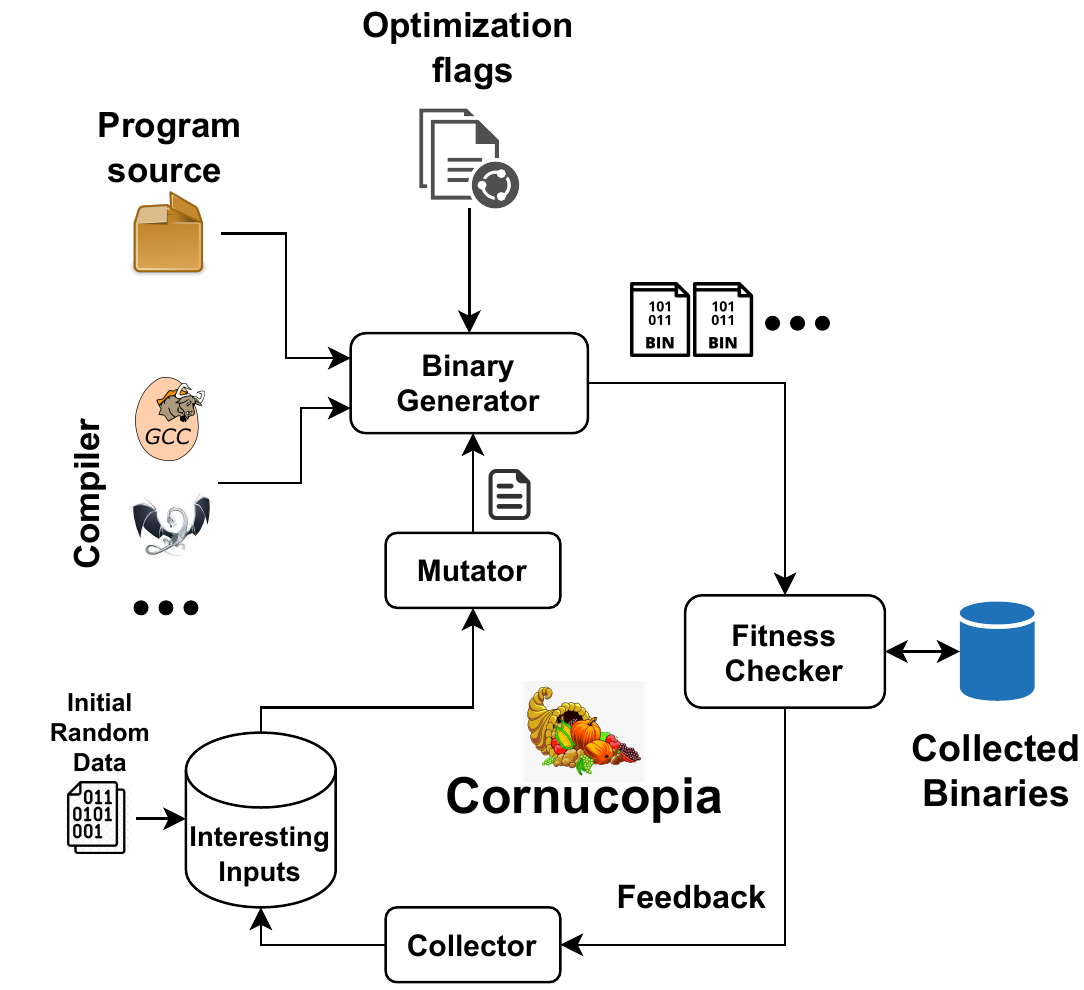}
\caption{Overview of~\ourtool.}
\label{fig:overview}
\end{figure}

This section presents an overview of~\ourtool{}, as shown in~\fig{fig:overview}.
The core technique of~\ourtool{} is the identification of the set of compiler flags
that affect the binary generated from the given source.
\ourtool{} starts with the program source package $S$, a compiler $C$,
the list of all flags $O$ supported by the compiler, and an
initial $|O|$ (\ie total number of flags) bytes of random data, used as an initial input. 

\ourtool{} uses feedback-guided mutation to select compiler flags that
have a high probability of changing the structure of the binary, as determined by a
configurable fitness function.
The~\emph{mutator} takes one of the interesting inputs (initially random data), mutates it, and
sends it to the~\emph{binary generator}.
The binary generator uses the data to select a certain subset of compiler flags $o_{i} \subseteq O$. (In other words, the input is used as a seed to select which compiler flags are used.)
These selected compiler flags are used to compile $S$ with $C$ to get a set of binaries $b$.
Note that each source package can result in multiple binaries.
For instance, compiling~\texttt{binutils.tar.gz} package results in 19 binaries,
such as~\texttt{objdump},~\texttt{nm}, etc.
The generated binaries $B$ are sent to a \emph{fitness checker}, which checks if these binaries are different from previously seen binaries and stores the newly-seen binaries, $B_{new} \subseteq B$, into a database.
\ourtool's fitness checker measures how~\emph{different}
the binaries in $B_{new}$ are from all the previously seen binaries from the same source package.
The measure of difference is converted into a floating-point number and sent as feedback to our collector.
The~\emph{collector} checks if the feedback value is greater than 0. If yes, it saves the corresponding input (generated by the mutator) into a weighted list of {\em interesting inputs} (inputs that yield differing binaries).

In the next iteration, the mutator again picks an input from the list of interesting inputs, such that the probability of picking an input is proportional to its feedback value. (This weighted sample means that inputs corresponding to compiler flags that created more varied binaries are preferred.)
The selected input is mutated and sent to the binary generator, and the process continues.
All the generated binaries will be saved into the database, and similar to the random testing process, the user can stop~\ourtool{} when she is satisfied with the generated binaries.

\section{Design}
\label{sec:Design}
The design of~\ourtool{} is built around the way fuzzing frameworks work, which expect to execute an ``input'' on a ``program'', generate an output, and from that output use a fitness function to decide whether or how to perturb the input. Crucially, in our setting, the ``program'' we are fuzzing is the combination of a compiler plus a program to be compiled (\eg LLVM plus {\tt objdump}). The {\em input} is the {\em compiler flags}. Section~\ref{subsec:bingen} describes how an input is mapped to compiler flags, and thence to generating an output. Section~\ref{subsec:fitnesschecker} describes how we design our fitness function.
\vspace{-4pt}
\subsection{Binary Generator}
\label{subsec:bingen}
The binary generator maps input bytes to compilation flags and uses these flags to compile a given source package to get a set of binaries.

\textbf{Mapping bytes to compiler flags:} 
For most of the flags, we map each input byte to a compiler flag.
The corresponding byte value indicates whether the option is selected or not.
However, directly using the byte value will result in unnecessary bias. 
For instance, consider that we enable an option by just checking whether the value of the byte is greater than 0. 
There is a 99\% (or 255/256) chance that the option is enabled, whereas there is only a 1\% (or 1/256) chance that the option will be disabled.
To avoid this bias, we use a modulus operation. Specifically, we compute~\textbf{byte\_value mod 2} and enable the flag if the resulting value is 1.
Similarly, for flags that expect a value from a fixed list, we use modulus to select a value uniformly from that list.
For instance, for~\textbf{--frame-pointer=<value>}, the~\textbf{<value>} can be either~\textbf{all},~\textbf{non-leaf}, or~\textbf{none}.  We use~\textbf{byte\_value mod 4} and enable the flag if the resulting value is greater than 0 and the~\textbf{<value>} can be either~\textbf{all},~\textbf{non-leaf}, or~\textbf{none} depending on whether the modulus result is 1 2 or 3 respectively.

For flags that take raw integers, we use 2 bytes, where the first byte (\textbf{mod 2}) indicates whether the option is enabled, and if enabled, the second byte is the value for the flag.
For instance, we map 2 bytes to the flag~\textbf{--stack-alignment=<uint>}.
The flag is selected when the~\textbf{first\_byte\_value mod 2} is 1 and the second byte is passed for~\textbf{<uint>},~\ie{}~\textbf{--stack-alignment=<second\_byte\_value>}.

We will ignore additional bytes if the input has more bytes than all the compiler flags. 
Similarly, we will not select the corresponding flags if the input has fewer bytes.

\textbf{Compiling using the selected flags: } 
%
We use a dynamic approach by hooking into the build process and
dynamically modifying every compiler invocation to include only the selected flags.
For instance, consider that our target compiler is~\textbf{clang} and
selected options are~\textbf{--addrsig} and~\textbf{--tailcallopt}. Our dynamic hook will replace every
compiler invocation, say~\textbf{gcc -O2 <source file(s)>}, with~\textbf{clang --addrsig --tailcallopt <source file(s)>}. We also inclu-
de all the preprocessor directives (\eg~\textbf{-D..}) and linker flags that were part
of the original compiler invocation.

\textbf{Handling conflicting flags: }
\label{subsub:handlingconflict}
The compiler flags can have constraints, including adverse interactions and dependency relationships.
Few flags can negatively influence each other, and turning them on together leads to a compilation error.
Some other flags may only work when another flag is specified.
For example,~\textbf{-ftree-slp-vectorize} may not have an effect when loop unrolling is disabled because SLP vectorizer may not have opportunities to vectorize the loop body if the loop is not unrolled. 

Automatically identifying conflicting compiler flags is a combinatorial problem~\ie requires enumerating all the possible flag combinations, which is intractable when there are a large number
of growing flags ($\sim$ 892 for x86 in~\textbf{clang-12.0}).
On the other hand, manually specifying conflicting flags for each compiler, as in~\bintuner{}, requires considerable effort.
We use a feedback-driven approach to handle this.
Specifically, if the compilation fails with selected flags, we compile using a default set of predefined flags (\eg \textbf{-O0}) and generate corresponding binaries.
Since the selection of any conflicting flags results in the same binary (\ie the one built with default flags), the fitness function (\sect{subsec:fitnesschecker}) will return a score of~\emph{zero} for these binaries.
The~\emph{zero} score will cause the corresponding input to be discarded by our collector, thereby steering~\ourtool{} away from generating inputs that result in conflicting compiler flags.
\vspace{-4pt}
\subsection{Fitness Checker}
\label{subsec:fitnesschecker}
The goal of the fitness checker is to compute how different the provided binaries are from all
the previously generated binaries from the same source package. We call this result~\ac{DScore}.

The score computation mechanism should be efficient. Otherwise, it will become a performance bottleneck, and then the overall cost will increase drastically.
Existing binary diffing techniques, such as BinDiff~\cite{dullien2005graph}, require disassembling the binary and performing lightweight analysis, increasing their execution time. For instance, BinDiff takes $\sim$5 min for a medium-sized binary.

There are well-known techniques in malware signature research that use heuristic methods to compute the similarity between two binaries.
We explore two such techniques and propose a custom difference score based on the percentage of unique functions.
In all of these techniques, the computed~\ac{DScore} is a floating-point number ranging from 0.0 to 1.0, where a larger value indicates a bigger difference.
As an optimization, before computing~\ac{DScore}, we check if the binary is not unique~\ie{}if we have already seen the exact binary, then we immediately return 0.
If the provided binaries are unique~\emph{\ie{}\ac{DScore} is greater than 0}, the fitness checker
also stores these binaries in a database.
We explore the following techniques to compute the~\ac{DScore} of a given binary.

\textbf{Piecewise Hashing: }
\label{subsub:piecewise}
Piecewise hashing or fuzzy hashing~\cite{kornblum2006identifying} is a well-known technique to compare binaries.
The comparison of fuzzy hashes results in a value ranging from 0.0 to 1.0 (the higher, the more different).
We explore two approaches to compute the~\ac{DScore} based on the piecewise hashing.

\textbf{Piecewise average (\pwavg)}:  Here, we compute the difference in the piecewise hash
of the given binary with all the previously seen binaries.
The final~\ac{DScore} is the average of all the hash difference scores.
The intuition behind the average is to compute a score that captures how different the current binary is when compared to~\emph{all} the previously seen binaries.

\textbf{Piecewise minimum (\pwmin)}: This technique is similar to the average one above. However, we select the minimum hash difference value instead of the average as the final~\ac{DScore}.
The intuition behind the minimum is to prioritize the generation of binaries that differ~\emph{largely} from all the previously seen binaries. 
If we consider the binary generation as a graph traversal, the average strategy can be considered as a Breadth-First traversal, whereas the minimum strategy is a Depth-First traversal.
We~\emph{do not} consider the maximum value because it unnecessarily prioritizes generating the same kind of binaries.
But, the goal of~\ourtool{} is to maximize the generation of different binaries.
For instance, consider a new binary $b$ with piecewise hash similarity of 0.9, 0.1, and 0.4 against binaries $x$, $y$ and, $z$, respectively.
Using the maximum value would return 0.9 as the~\ac{DScore}, thus maximizing the generation of binaries similar to $b$.
However, the hash difference value 0.1 indicates that binary $b$ is very similar to $y$.
Hence, using the maximum value may unnecessarily prioritize the generation of similar binaries and decrease the overall variety of binaries.

\textbf{Normalized Compression Distance: }
\label{subsub:ncd}
\ac{NCD} is another well-known technique to compute difference based on an information-theoretic measure~\cite{alshahwan2020detecting}.
Specifically,~\ac{NCD} infers the degree of similarity between arbitrary byte
sequences by the amount of space saved after compression.
Previous works~\cite{ren2021unleashing} which use~\ac{NCD} have shown to be effective at capturing the difference between two arbitrary byte sequences. \ac{NCD} score ranges from 0.0 to 1.0 (the higher, the more different). Similar to Piecewise hashing (\sect{subsub:piecewise}), we define \textbf{\ac{NCD} average (\ncdavg)} and \textbf{\ac{NCD} minimum (\ncdmin)}. 

\textbf{Percentage of Unique Functions (\funchash): }
\label{subsub:functionhash}
Here, we compute the difference score as the percentage of unique functions in the provided binary.
We determine unique functions as follows:
For each function, we compute function hash, which is the hash of the binary code of the function.
We use this function hash to see if any previously seen binaries have a function with the same hash. If not, the function is considered unique.
Finally, the~\ac{DScore} is computed as the percentage of unique functions over the total number of functions in the binary.
The intuition here is to use function level similarity rather than byte-sequences based similarity techniques as used in the previous two approaches.
\subsection{Collector and Mutator}
\label{subsec:collectorandmutator}
The collector receives the feedback (\ie{}\ac{DScore}) for each input and stores the input
in a weighted list according to the value of the score.
The collector discards inputs with a feedback score of 0.
The weighted list is organized such that the probability of selecting an element from the list
is proportional to its feedback score.

The mutator selects one or more inputs from the weighted list and performs various mutations on the bytes of the inputs.
We use mutation strategies, such as~\emph{bit flips},~\emph{byte flips}, and~\emph{splicing}, that are shown to be effective in fuzz testing~\cite{fuzzingmutation}.

\ifnotreport
Refer our extended report~\cite{llvmrandomextended} for implementation details.
\fi

\ifreport
\vspace{-2pt}
\section{Implementation}
\label{sec:implementation}
\ifnotreport
We modified~\AFL{} to implement~\emph{collector} and~\emph{mutator}.
This simplified our implementation, as~\AFL{} already has various mutation strategies, achieves very high execution rates, and also has inbuilt parallelism support.

The binary generator is implemented as a C++ program that uses a modified version of~\ac{WLLVM}~\cite{wllvmtool}.

The fitness checker is implemented as a RESTful server (\fitserver) that encapsulates the computation of the~\ac{DScore}.
The \fitserver exposes a REST interface, which accepts a binary and returns the difference score by using one of the configured techniques (\sect{subsec:fitnesschecker}).
We present more details of the implementation in our extended report~\cite{llvmrandomextended}.

\else
In this section, we present the implementation details of various components of~\ourtool{}.
Our implementation is agnostic to the compiler and~\ac{ISA} of the target binary.
Furthermore, our system is modular and extensible.

\textbf{\AFL{} modifications: }
We modified~\AFL{} to implement~\emph{collector} and~\emph{mutator}.
This simplified our implementation, as~\AFL{} already has various mutation strategies and uses techniques, such as fork server, to achieve very high execution rates.
We also configured the master/slave mode of~\AFL{} to implement the parallel mode for \ourtool{}, where multiple instances generate binaries for a given program by sharing interesting inputs. 
We implemented our collector by modifying~\AFL{}'s coverage computation logic.
Specifically, instead of using the coverage bitmap~\cite{coveragemap}, we modify~\AFL{} to use the~\ac{DScore} to determine whether the input is interesting.
We use a predefined region in the coverage shared memory~\cite{coveragemap} to communicate the~\ac{DScore}.
All interesting inputs will be stored in a weighted list sorted according to their~\ac{DScore}.

\textbf{Binary generator as custom mutator: }
The important component of the binary generator is hooking into the build system,
which we implement by modifying~\ac{WLLVM}~\cite{wllvmtool}, which provides python-based compiler wrappers. 
These wrappers provide programmable hooks to modify all compiler invocations.
We implemented our binary generator as a C++ program that maps a given sequence of bytes
to a set of compiler flags and selects the appropriate flags according to the logic as explained in~\sect{subsec:bingen}.
Next, we use these selected flags to invoke our modified~\ac{WLLVM}, which will
modify all compiler invocations to use the selected flags.
If any of the compiler invocations fail because of conflicting options (\sect{subsub:handlingconflict}), we rerun the compilation using a default set of predefined flags (\ie \textbf{-O0}).

We use the custom mutator support~\cite{aflcustommutator} of~\AFL{} to integrate our binary generator.
Specifically, every input generated by~\AFL{} will be sent to our binary generator (\ie custom mutator).
Our binary generator, as mentioned above, maps the input bytes to appropriate compiler flags
and generates a binary using these flags. 
The generated binary will be returned to~\AFL{}, which will forward it to our fitness checker.

\textbf{Optimization for LLVM: }
We have an optimized mode for LLVM, where we use bitcode files generated using O0  (\ie no optimizations) instead of compiling from source code every time.
We run the optimizations using~\ac{llc}, a bitcode compiler.
We do an initial pass for a given source package of getting bitcode files (instead of binaries) using O0.
Subsequent calls to the binary generator will pass the selected flags to run~\ac{llc} using bitcode files.
As we show in~\sect{subsub:bingeneffectiveness}, this greatly improves the overall yield as we avoid the unnecessarily step of going through the compiler frontend every time for the same source package.
For other compilers that do not have an externally accessible intermediate representation, such as~\textbf{gcc}, we take the usual route of running the whole compiler using~\ac{WLLVM}.

\textbf{Fitness checker as a service: }
We split the fitness checker into two components (i) Proxy program (\proxyp) and (ii) Fitness checker server (\fitserver).
Splitting the fitness checker provides flexibility in developing custom fitness checking functions. Users interested in developing custom fitness checking functions need not worry about the internals of~\ourtool{} and just need to expose an interface that accepts a binary and returns a difference score.

The proxy program (\proxyp) is written in C and is the main program to which~\AFL{} passes the binary (provided by our binary generator).
The Fitness checker server (\fitserver) is a server program that encapsulates the computation of the~\ac{DScore}.
The \fitserver exposes a REST interface, which accepts a binary and returns the difference score by using one of the configured techniques (\sect{subsec:fitnesschecker}).
It also stores the provided binary (if unique) into a database.
We only~\emph{focus on the changes in the code (\ie \textbf{.text} section) region of the binary}.
In other words, we consider two binaries (\ie ELF files) as different~\emph{if and only if} they differ in their~\textbf{.text} section.

For piecewise hashing, we use~\texttt{ssdeep} python module.
For~\ac{NCD}, we use lzma~\cite{gilbert1992lempel} as the underlying compression algorithm, as it is shown to be a good candidate~\cite{borbely2016normalized}.

We start our modified~\AFL{} by passing~\proxyp{} as the target program to be fuzz tested.
Our process starts by~\AFL{} sending the generated binary to~\proxyp{}, which sends the binary to~\fitserver{} and receives a~\ac{DScore}.
The~\proxyp{} will store the~\ac{DScore} into the preconfigured region of the shared memory and exits.
As mentioned above,~\AFL{} retrieves the~\ac{DScore} from the shared memory and continues generating its next input.

\textbf{Extensibility: }
Our implementation of~\ourtool{} is extensible and can be easily configured to use a new compiler, source package, and custom fitness checking functions. We provide the exact details in~\apdx{apdx:extensibility}.
\fi

\fi
\vspace{-2pt}
\section{Evaluation}
\label{sec:evaluation}
We evaluate~\ourtool{} to demonstrate its effectiveness in generating binaries and their ability to test the robustness of various binary analysis tools.
We pose the following research questions to guide our evaluation:

\noindent
\textbf{RQ1: Effectiveness:} How effective is~\ourtool{} in generating binaries, and how do different fitness metrics affect the quality and quantity of the generated binaries?

\noindent
\textbf{RQ2:~\ourtool{} vs.~\bintuner{}:} How effective is~\ourtool{} compared to~\bintuner{}, a recent approach that also uses compiler flags to generate binaries?


\noindent
\textbf{RQ3: Applicability to test static analysis tools:} How effective is the dataset generated by~\ourtool{} in testing binary static analysis tools?

\noindent
\textbf{RQ4: Applicability to test~\ac{ML} tools:} How effective is the dataset generated by~\ourtool{} in tesing~\ac{ML} tools?

\subsection{Setup}
\subsubsection{Dataset and Compiler}
\label{subsec:datasetcomp}
We choose \textbf{clang} (or LLVM) version 12 as our target compiler, which is the latest and most stable version available during our experimentation.
Our binary
generator for~\clangversion{} uses pre-generated LLVM Bitcode files as an
optimization to avoid rerunning frontend
for the same sources.

We collected source packages by scrapping official Debian package repositories, compiled them, and randomly selected~\numllvmpackages{} bitcode files for each of the four popular architectures,\ie\xeightsix,~\xsixfour,~\arm, and~\mips. 
We will refer to individual binaries or bitcode files as programs.
Table~\ref{tab:variationsarchclang} shows the number of programs selected and available optimization flags in~\clangversion{} for each architecture.
Note that the number of programs is limited by resource constraints; specifically, the availability of machines at our disposal.

\subsubsection{Machine Setup and Runtime}
\label{subsec:fuzztime}
\label{subsec:compcrash}
We used a server with Intel Xeon 5215 CPU and ran~\ourtool{} on each program for~\fuzztime{} hours. We ensured that each program ran on a processor core and avoided overloading the server.

\subsection{Effectiveness}
\label{subsec:evaleffectiveness}
As explained in~\sect{subsec:fitnesschecker}, there are various lightweight approaches to compute the difference score that can guide our mutations.
There is also another approach, as suggested by a recent work~\cite{ren2021unleashing} where they
take the~\ac{NCD} score of the binary with the binary
compiled with \textbf{-O0} as the difference score, which we denote as~\ncdo.
First, we will evaluate the relative effectiveness of our approaches \pwavg,~\pwmin,~\ncdavg,~\ncdmin, and,~\funchash along with~\ncdo.

\subsubsection{Effectiveness of different computation approaches}
\label{subsub:impactdiff}
\vspace{-3pt}
We choose three programs of different sizes~\texttt{eot2ttf} (5.5K, small),~\texttt{lscpu} (270K, medium), and,~\texttt{nab\_r} (1.1M, large) for this experiment.
For each of these programs, we ran~\ourtool{} with different approaches for six hours each. 
In summary, we had 18 (6 approaches * 3 programs) variations, with each running for six hours.
To normalize the effects of randomness, we repeated the whole experiment eight times.
We found that our function hash mechanism (\funchash) resulted in the largest number of unique binaries generated for all three programs for most of the iterations.
The second best technique is~\ac{PIECEWISE} minimum (\pwmin).

To compare the quality of the generated binaries, we computed the~\ac{NCD} score of each binary against the non-optimized \ie \textbf{-O0} compiled) binary.
We found that, on average, the binaries generated by \funchash have the highest difference score (\ie~\emph{more different variants}) of \textbf{\textit{0.79}} compared to all the other fitness functions. 

\emph{This shows that our \funchash technique to compute difference score is both quantitatively (\ie more unique binaries in a fixed interval of time) and qualitatively (\ie more different binaries) more effective at generating unique binaries when used with~\ourtool{}.}

There are two main reasons for the improved effectiveness of \funchash: 
(i) Most of the optimizations in compilers are intraprocedural and work independently on each function. 
(ii) Functions within a program share similar characteristics~\cite{meng2021bran, nguyen2013study}. For instance, most of the functions in a string processing library work on strings~\ie \textbf{char *} type variables.
Hence optimization flags that affect a function in a program most likely also affect other functions in the same program as these functions share similar characteristics.
Our \funchash approach exploits this by assigning a higher score to the flag combinations that affect more functions in the program.

We also ran the experiment by avoiding the precise difference score but rather using a 1/0 binary feedback,~\ie whether the generated binary is different (1) or not (0). We observed that all approaches suffered and generated fewer binaries compared to the precise difference score versions. This indicates that using a precise difference score is important for generating large number of unique binaries.
The potential reason is that using a precise score helps in guiding the search towards more productive flag combinations while 1/0 will do a random search.
\vspace{-6pt}
\subsubsection{Binary Generation Effectiveness}
\label{subsub:bingeneffectiveness}
We use the most effective difference score approach,\ie function hash (\funchash), to evaluate the overall effectiveness of~\ourtool.
As mentioned in~\sect{subsec:fuzztime}, we ran~\ourtool for~\fuzztime{} hours for each program-architecture combination.
The summary of the results is shown in~\tbl{tab:variationsarchclang}.
In total~\ourtool generated~\totalclangbinaries unique variants across four architectures for~\numllvmpackages programs, with an average of~\avgclangbinnum and median of~\medianclangbinnumber variants per program across all the architectures (The fine-grained split is discussed in 
\ifreport
\fig{fig:varclangtotal} of Appendix
\fi
).

\noindent
\textbf{Variants across each architecture:}
Interestingly, as shown in~\tbl{tab:variationsarchclang} the number of generated binaries differs across architectures.
Specifically, there are $\sim$15\% more binaries in ARM and MIPS, which have a \ac{RISC}~\ac{ISA}, compared to x86 and x64, having a~\ac{CISC}~\ac{ISA}.

\begin{figure}[t]
\includegraphics[width=\columnwidth]{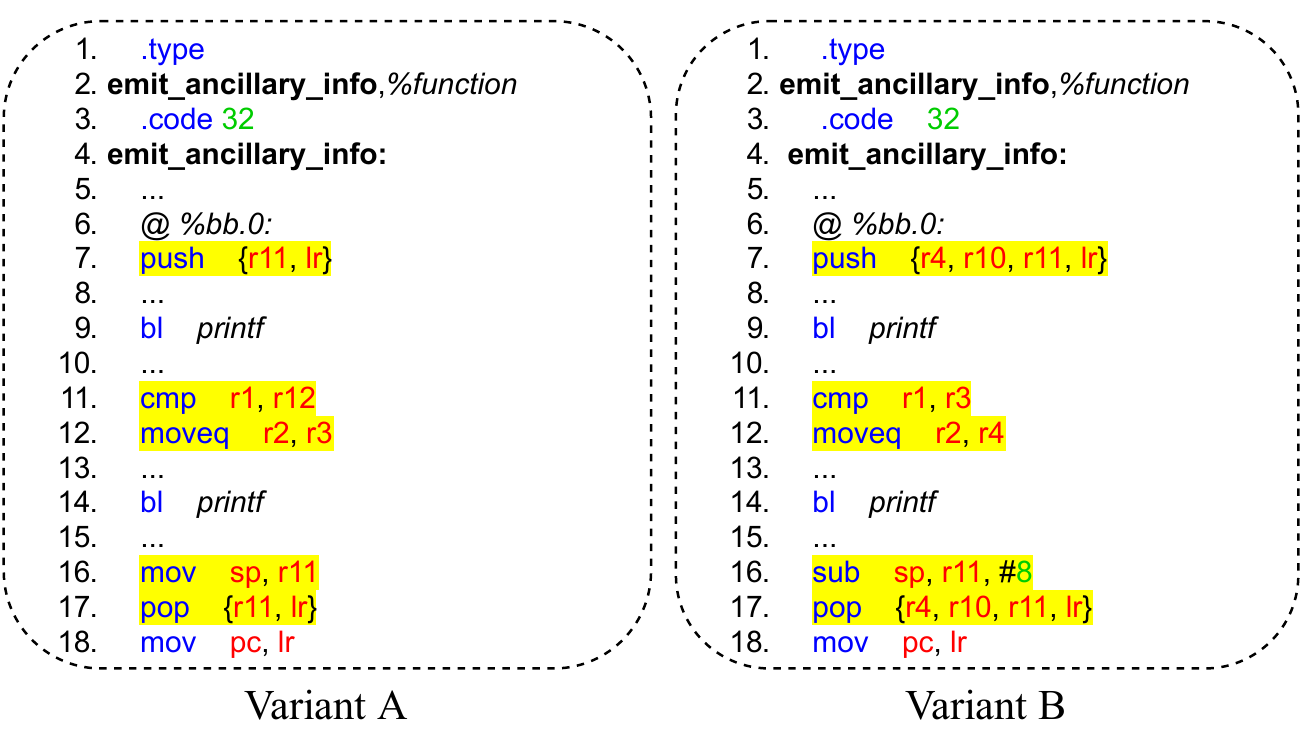}
\caption{Figure showing ARM assembly for 2 different variants generated from source program \textbf{cat}. Variant A uses registers \textbf{(r12, r3)} in place of \textbf{(r3, r4)} (Variant B) for the same function.}
\label{fig:registerDiff}
\end{figure}

The main reason for this is the difference in the underlying~\ac{ISA} and corresponding optimization opportunities.
There are more general-purpose registers in ARM and MIPS than x86 and x64, which increases the compiler's choices for register allocation.
An example illustration is in one of our binaries as shown in~\fig{fig:registerDiff}, here compiler choose \textbf{r12} and \textbf{r3} in the left version v/s \textbf{r3} and \textbf{r4} in the right version, this further caused register spill (line 7 and 17) to occur in the right version.
Furthermore, the fixed-length instructions in ARM and MIPS results in relatively dense basic blocks,~\ie the average number of~\emph{instructions} in a basic block are more than in x86 and x64~\cite{blem2015isa}. This further increases optimization opportunities.

We evaluated~\ourtool{} on other aspects and presented the results in our extended report~\cite{llvmrandomextended}.
Our results show that~\ourtool{} is effective at generating a large number of different binaries and can explore the variants that are not covered by the standard optimization levels~\ie O0, O1, O2, and O3.

\ifreport

\noindent
\textbf{Size of programs v/s number of binaries generated:}
We observed that the number of generated binaries follows a gaussian distribution w.r.t the program size. This makes sense as larger programs have more optimization opportunities, hence generating more binaries. However, compilation time also increases with program size, which limits the number of iterations and consequently results in fewer binaries.
We present a detailed analysis along with corresponding results in
\ifreport
~\apdx{appendix:sizevsnumber}.
\else
in our extended report~\cite{llvmrandomextended}
\fi
.

%

\noindent
\textbf{Quality of the generated binaries:}
In addition to the many (quantity) binaries, we also want~\ourtool to generate different varieties of binaries (quality).
We use~\textit{Inverse Bindiff} (\bindiff) scores (\ie 1 - similarity (bindiff) score) to measure the difference between two binaries. 
We performed a detailed analysis of the binaries generated by~\ourtool{} against those generated by standard optimization levels (\ie O1, O2, and, O3).
A detailed analysis of the results is in
\ifreport
~\apdx{sec:qualgenbin}.
\else
our extended report~\cite{llvmrandomextended}.
\fi
Our results show that~\ourtool{} is effective at generating a large number of different binaries and can explore the variants that are not covered by the standard optimization levels~\ie O0, O1, O2, and O3.
\fi
%
%
%
\begin{table}[]
\centering
\resizebox{\columnwidth}{!}{
\begin{tabular}{c|r|r}
\toprule
\textbf{Arch (Available Flags)} & \multicolumn{1}{l|}{\textbf{Binaries}} & \multicolumn{1}{l}{\textbf{Avg. Binaries Per Program}} \\ 
\midrule
\rowcolor{black!15} {\xeightsix (892)}  &   63,197  & \multicolumn{1}{r}{330.87}   \\  
{\xsixfour (892)}                       &   74,169  & \multicolumn{1}{r}{388.32}   \\ 
\rowcolor{black!15} {\arm (876)}        &   83,701  & \multicolumn{1}{r}{438.23}   \\ 
{\mips (866)}                           &   87,192  & \multicolumn{1}{r}{456.50}   \\ 
\midrule\textbf{Grand Total}            &   308,269 &   \multicolumn{1}{c}{N/A}                                                                 \\ 
\bottomrule
\end{tabular}
}
\caption{Performance of~\ourtool{}: The number of binaries generated for each architecture for 191 programs. Each program-architecture combination is run for~\fuzztime hours. The number in the parenthesis show the total number of available optimization flags for that architecture.}
\label{tab:variationsarchclang}
\end{table}

\ifreport
\noindent
\textbf{Scaling~\ourtool{} with parallel mode:}
\ifnotreport
As mentioned in~\sect{sec:implementation}, we engineered~\ourtool{} to run in parallel mode, where multiple instances try to generate binaries for a given program by sharing interesting test cases across all the instances.
For each program, parallel mode generated 5X-41X (with an average of 21X) more binaries than single instance mode, demonstrating the effectiveness of parallel mode and scalability of~\ourtool{} with more processor cores.
We present the details our this experiment in our extended report~\cite{llvmrandomextended}.
\else
As mentioned in~\sect{sec:implementation}, we engineered~\ourtool{} to run in parallel mode, where multiple instances try to generate binaries for a given program by sharing interesting test cases across all the instances.
To evaluate this, we tested~\ourtool{} in parallel mode with six instances on six programs of varying sizes (5.5KB - 209KB).
We ran parallel mode and single instance mode for six hours per program.
\fig{fig:afl-parallel} in~\apdx{apdx:parellmodecornu} shows the number of unique binaries generated by both of these modes across different programs.
For each program, parallel mode generated 5X-41X (with an average of 21X) more binaries than single instance mode, demonstrating the effectiveness of parallel mode and scalability of~\ourtool{} with more processor cores.
\fi
\fi

\subsubsection{Compiler Crashes}
\label{subsubsec:compilercrashes}
Although unintended,~\ourtool{} could be used to test optimization schedulers in compilers.
As explained in~\sect{subsec:bingen}, our binary generator repeatedly invokes the compiler with different combinations of optimization flags on various programs.
Consequently, while generating binaries for different programs,~\ourtool{} is essentially testing optimization schedulers, although in a blackbox manner.
Nonetheless, in our experiments,~\emph{we found approx.~\numllvmcrashes crashes (\ie \textbf{segfaults}) in the optimization scheduler of~\clangversion}.
An example of one such crash is shown in
\ifreport
~\lst{lst:llvmcompcrash} (Appendix)
\fi
.
We analyzed one of these crashes and identified that the  \textbf{--pre-RA-sched=vliw-td} optimization flag is the root cause.
This is not a trivial issue to find because triggering the crash requires specific program structure.
We reported all our crashes and have been acknowledged by the LLVM team as real bugs.
They are currently working on fixing these bugs.

\ifnotreport
We also extended~\ourtool{} with~\gccversion{} and presented its results in our extended report~\cite{llvmrandomextended}.
\else
\vspace{-4pt}
\subsubsection{Extending to Different Compilers}
\ourtool{} is extensible in accommodating different compilers. We present the results of extending~\ourtool{} with~\gccversion{} in
\ifreport
~\apdx{apdx:differentcompilers}.
\fi
\fi













\subsection{\ourtool{} vs.~\bintuner{}}
\label{subsub:cornucopavsbintuner}
As mentioned in~\sect{sec:introduction},~\bintuner{} uses a search-based iterative compilation (based on OpenTuner~\cite{ansel2014opentuner}) to find optimization sequences that can maximize the amount of binary code differences.
~\bintuner{} requires an explicit specification of conflicting compiler flags in the form of first-order logic formulas, which requires an in-depth understanding of the flags. This process can be tedious, especially when we need to do this for every architecture supported by the compiler (\ie{}\xeightsix,~\xsixfour,~\arm,~\mips, etc) and for all desired compiler versions.
This imposes considerable~\emph{overhead for binary analysis tool developers to use~\bintuner{}}.
Furthermore, the implementation of~\emph{\bintuner{} does not support parallelism}, and as such,~\bintuner{} cannot be used in a multi-processor/multi-threaded manner to improve its throughput.

However,~\tool{} only requires specifying the compiler and a corresponding list of supported optimization flags. It does not require~\emph{specifying conflicting flags}. Our feedback-driven mechanism (\sect{subsec:bingen}) enables~\tool to automatically steer away from using conflicting flags.
The modular design of~\tool enables it to be trivially parallelizable by using multiple mutators, all sharing the same interesting inputs source.
As shown in~\sect{subsec:evaleffectiveness}, running~\tool in parallel mode with six instances resulted in an average of 21X more binaries.

To have an analytical comparison, we perform the following two experiments on the programs on which~\bintuner was evaluated.
Specifically, we use SPECint 2006, Coreutils, and OpenSSL.
\subsubsection{\tool{} with~\bintuner{}'s fitness function (\cornwithbin{})}
\label{subsub:cornuwithbinfitness}
In this first experiment, we evaluate the binary generation effectiveness of~\bintuner{}'s fitness function when used in~\tool{}.
Specifically, as in~\bintuner{}, we use~\ac{NCD} score of the generated binary with its \textbf{-O0} version as the feedback (\ie{}\ac{DScore}) for the collector in~\tool{}, denoted as~\cornwithbin{}.

On average~\cornwithbin{} generated~\avgcorcbbinaries{} binaries vs~\avgcorbinaries{} generated by~\ourtool with the function hash score (\funchash).
The~\fig{fig:numberbinariesbintuner} shows the results across all the programs (Note that the y-axis is in logarithmic (base 10) scale). Except for~\texttt{447.dealII} and ~\texttt{483.xalancbmk},~\ourtool generated a large number of binaries, specifically, $\sim$7X more than~\cornwithbin{}.
The low yield in~\texttt{447.dealII} and ~\texttt{483.xalancbmk} is because of their large size and the randomness in mutation techniques having less time to explore other effective optimization flag combinations.
The reason for the increased effectivenss of~\ourtool is because~\bintuner{}'s fitness function (\ac{NCD} with  \textbf{-O0}) maximizes the generation of a highly different binary rather than generating a large number of diverse binaries.
For instance,~\cornwithbin{} likely will not generate highly different binaries that have the same~\ac{NCD} score with  \textbf{-O0}.
\subsubsection{Binary Generation Effectiveness}
\label{subsub:bingeneffectiveness}
\begin{figure*}[h]
\centering
\includegraphics[scale=0.35]{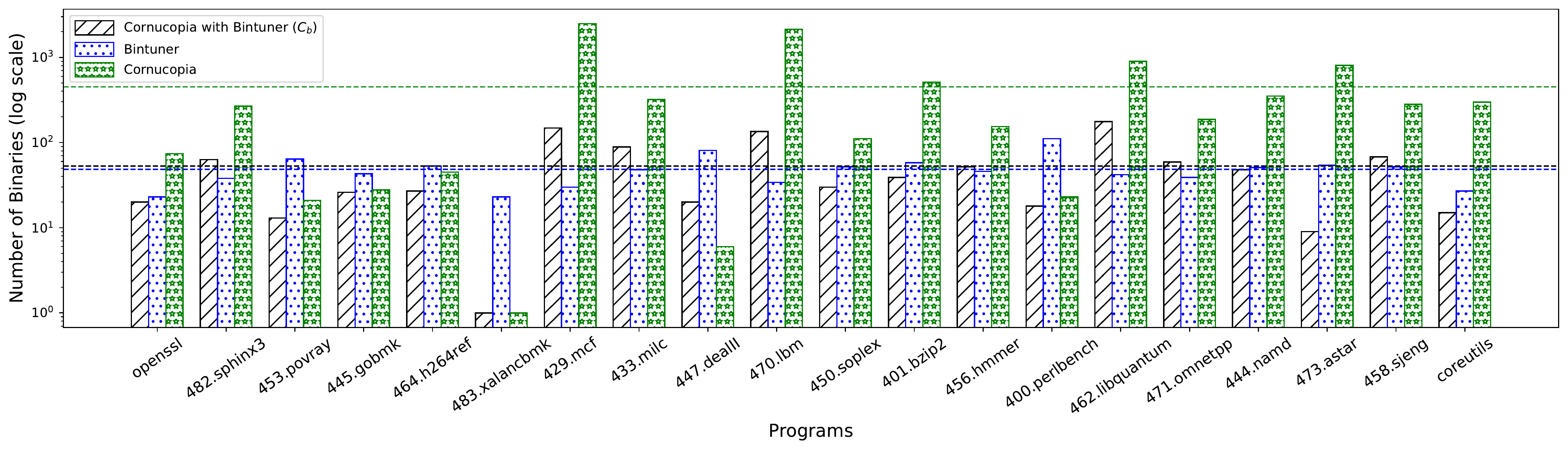}
\caption{Number of binaries generated by each technique for OpenSSL, SPEC 2006, and Coreutils programs. The horizontal lines represent average number of binaries per-program generated by each technique.}
\label{fig:numberbinariesbintuner}
\end{figure*}

For this experiment, we ran~\ourtool for 6 hours and~\bintuner until it converges or 6 hours (whichever is the latest).
On average~\bintuner generated~\avgbintunerbinaries{} binaries vs~\avgcorbinaries{} generated by~\ourtool with the function hash score (\funchash), with~\fig{fig:numberbinariesbintuner} showing the results across all the programs.
Except for five programs,~\ourtool was able to generate more binaries ($\sim$8X on average) than~\bintuner{}.
The low yield for a few programs is because of their large size and~\ourtool getting less number of iterations in identifying the optimization flags that are effective for these binaries. 
However,~\bintuner, based on OpenTuner~\cite{ansel2014opentuner}, uses more systematic exploratory techniques and can quickly identify the potent optimization flags.
For instance, the bitcode file for~\texttt{483.xalancbmk} is 13MB in size, and compilation of it takes $\sim$ 6 minutes. Consequently,~\ourtool gets less time to explore different flag combinations and learn which flags are effective.
We confirmed this by running~\ourtool{} in parallel mode with six cores and observed that we got considerably more binaries than~\bintuner.
\subsubsection{Quality of the Generated Binaries}
\label{subsub:bingenquality}
We used BinDiff scores to evaluate the quality of binaries generated by different techniques (\bintuner,~\cornwithbin{}, and~\ourtool) and~\fig{fig:bindiffbintuner} shows the cumulative distributive function (CDF) of the scores across all binaries generated for all programs by each of the corresponding techniques.~\footnote{A point (x, y) on a line indicates y\% of the binaries have their BinDiff score less than or equal to x.}
The score ranges from 0 to 1, and it indicates the amount of difference (\ie larger the score higher the difference).
First, as expected,~\bintuner{} was able to generate binaries with the largest difference ($\sim$0.95) against its  \textbf{-O0} and  \textbf{-O3} versions. However, its steeper curve shows little variance, i.e., most of the~\bintuner generated binaries are similar and have high diffence against its  \textbf{-O0} and  \textbf{-O3} versions.
The less steep curves of~\ourtool and~\cornwithbin{} show that they were able to generate more varied binaries, albeit with a lower difference ($\sim$0.45) against its  \textbf{-O0} and  \textbf{-O3} versions.

We also compared the best binary (\ie with the highest BinDiff score) generated by~\bintuner with the binaries generated by~\ourtool.
The~\fig{fig:bindifftunercornu} shows the CDF of the corresponding score. The steeper curve towards the right indicates that~\emph{most of the~\ourtool generated binaries are quite different from those of~\bintuner{}'s}. Specifically, 50\% of the binaries have their BinDiff scores between 0.75-0.95. 
This shows that~\ourtool is exploring the binary generation space different from that of~\bintuner.
In summary,~\bintuner is effective at generating binaries highly different from its  \textbf{-O0}/ \textbf{-O3} version, but the generated binaries have less variance.
 However,~\ourtool is a complementary approach and can efficiently generate a large number of binaries with relatively high variance by exploring different binary generation spaces.
\definecolor{avgcolor}{rgb}{0.59, 0.44, 0.84}
\definecolor{mediancolor}{rgb}{0.34, 0.01, 0.1}
\definecolor{maxcolor}{rgb}{0.0, 0.0, 0.61}

\definecolor{O0vsO1}{rgb}{0.55, 0.71, 0.0}
\definecolor{O0vsO2}{rgb}{1.0, 0.77, 0.05}
\definecolor{O0vsO3}{rgb}{0.89, 0.35, 0.13}

\begin{figure*}[h]
\begin{minipage}{0.3\textwidth}
\centering
\resizebox{0.775\textwidth}{!}{
\begin{tikzpicture}
\begin{axis}[
	xlabel=BinDiff score,
	ylabel=Percentage of Binaries,
	ytick style={font=\footnotesize},
	xtick style={font=\footnotesize},
	ymin=0,
	ylabel near ticks,
	xlabel near ticks,
	ymax=100,
	xtick={0, 0.05, 0.1, 0.15, 0.2, 0.25, 0.3, 0.35, 0.4, 0.45, 0.5, 0.55, 0.6, 0.65, 0.7, 0.75, 0.8, 0.85, 0.9, 0.95, 1.0},
	ytick={10,20,...,100},
    yticklabel style={font=\footnotesize},
	xticklabel style={font=\footnotesize, rotate=60},
	xmin=0,
	xmax=1.0,
	grid=both,
    legend style={at={(0.5,-0.30)},anchor=north, font=\tiny},
    legend columns=3,
    ]

\addplot[color=avgcolor,mark=square*] table[col sep=comma] {data/bindiff/bintuner/bintunerO0.txt};\addlegendentry{\bintuner{}};
\addplot[color=mediancolor,mark=triangle*] table[col sep=comma] {data/bindiff/bintuner/ccO0.txt};\addlegendentry{\ourtool{}};
\addplot[color=maxcolor,mark=*] table[col sep=comma] {data/bindiff/bintuner/ccbO0.txt};\addlegendentry{\cornwithbin{}};

\end{axis}
\end{tikzpicture}
}
\caption{Against O0 Binary}
\label{fig:bindiffxeightsix}
\end{minipage}%
\begin{minipage}{0.3\textwidth}
 \centering
  \resizebox{0.7\textwidth}{!}{
    \begin{tikzpicture}
\begin{axis}[
	xlabel=BinDiff score,
	yticklabels={,,},
	xtick style={font=\footnotesize},
	ymin=0,
	ylabel near ticks,
	xlabel near ticks,
	ymax=100,
	xtick={0, 0.05, 0.1, 0.15, 0.2, 0.25, 0.3, 0.35, 0.4, 0.45, 0.5, 0.55, 0.6, 0.65, 0.7, 0.75, 0.8, 0.85, 0.9, 0.95, 1.0},
	ytick={10,20,...,100},
    yticklabel style={font=\footnotesize},
	xticklabel style={font=\footnotesize, rotate=60},
	xmin=0,
	xmax=1.0,
	grid=both,
    legend style={at={(0.5,-0.30)},anchor=north, font=\tiny},
    legend columns=3,
    ]

\addplot[color=avgcolor,mark=square*] table[col sep=comma] {data/bindiff/bintuner/bintunerO3.txt};\addlegendentry{\bintuner{}};
\addplot[color=mediancolor,mark=triangle*] table[col sep=comma] {data/bindiff/bintuner/ccO3.txt};\addlegendentry{\ourtool{}};
\addplot[color=maxcolor,mark=*] table[col sep=comma] {data/bindiff/bintuner/ccbO3.txt};\addlegendentry{\cornwithbin{}};

\end{axis}
\end{tikzpicture}
    }
    \caption{Against O3 Binary}
    \label{fig:bindiffxsixfour}

\end{minipage}%
\begin{minipage}{0.3\textwidth}

  \resizebox{0.7\textwidth}{!}{
    \begin{tikzpicture}
\begin{axis}[
	xlabel=BinDiff score,
	yticklabels={,,},
	xtick style={font=\footnotesize},
	ymin=0,
	ylabel near ticks,
	xlabel near ticks,
	ymax=100,
	xtick={0, 0.05, 0.1, 0.15, 0.2, 0.25, 0.3, 0.35, 0.4, 0.45, 0.5, 0.55, 0.6, 0.65, 0.7, 0.75, 0.8, 0.85, 0.9, 0.95, 1.0},
	ytick={10,20,...,100},
    yticklabel style={font=\footnotesize},
	xticklabel style={font=\footnotesize, rotate=60},
	xmin=0,
	xmax=1.0,
	grid=both,
    legend style={at={(0.5,-0.3)},anchor=north, font=\tiny},
    legend columns=3,
    ]

\addplot[color=mediancolor,mark=triangle*] table[col sep=comma] {data/bindiff/bintuner/bintunercor.txt};\addlegendentry{\ourtool{}};

\end{axis}
\end{tikzpicture}
   }
    \caption{Against best BinTuner binary}
    \label{fig:bindifftunercornu}
\end{minipage}%
\caption{CDF of~\textit{Bindiff} scores of generated binaries by each of the techniques against O0, O3 and best bintuner binary.}
\label{fig:bindiffbintuner}

\end{figure*}
\subsection{Applicability to Test Static Analysis Tools}
\label{subsub:robuststaticanalysis}
We used four popular binary static analysis tools,~\ie Free and open source:~\angr{},~\ghidra{}, and ~\radare{}; Commercial:~\idapro{} to evaluate the effectiveness of \ourtool{} generated binaries in testing these tools.
We choose analyses that are supported by all these tools.
Specifically, we choose the following:

\noindent
\textbf{Function Boundary Detection (FBD)~\cite{bao2014byteweight}:}
\label{subsub:funcbound}
This analysis generates a set of function boundaries, where each boundary is a pair of addresses indicating the address of the first and last instruction of a function.
We got the ground truth information for FBD from debug information~\cite{dwarfformat} of binaries, specifically, the symbol table~\cite{youngdale1995kernel}.
%
%

\noindent
\textbf{Calling Convention Recovery (CCR):} This analysis aims to find the signature~\cite{lin2021function} of all functions in the binary.
For our experiment, we only consider the~\emph{number} of parameters.
Like FBD, we got the CCR ground truth for each binary using the debug information embedded in it.

To test these two analyses, we compare the ground truth of each binary with the results produced by each tool.
For each analysis, we assigned a fixed time of 24 hours for each architecture, randomly picked binaries, and tested them with each tool with a timeout of 10 minutes - most of the tools were able to complete within the timeout except for~\angr, which timed out for a relatively few large binaries.

~\tbl{tab:tooldifftesting} shows the result across the selected tools.
Here,~\failures indicate the number of binaries with single tool failures,~\ie only the corresponding tool failed.
\failurem indicate multi-tool failures,~\ie two or more tools failed.
Finally~\successa indicates binaries where all tools succeeded,~\ie all tools correctly identified function boundaries for these binaries.

For FBD (Top part of~\tbl{tab:tooldifftesting}), on average, all tools correctly identified boundaries for only 19.97\% of the binaries across all architectures.
Unfortunately, \emph{none} of the tools correctly identified function boundaries for 42.15\% of the binaries as indicated by the last row of \failurem column.
For instance, for a binary of  \textbf{fallocate} compiled for \mips, with 172 functions, all the tools except \ghidra{} failed to precisely detect all the functions. 
\angr{} overestimated and detected 182 functions, whereas \idapro{} and \radare{} missed several functions and detected 158 and 100 functions, respectively.
\radare performs worst by failing on most binaries across all architectures.
\angr performed relatively well on \xeightsix and~\xsixfour, confirming previous studies~\cite{Pang2021SoKAY}.
However, across all architectures,~\idapro performs better on average.
For function boundary detection,~\angr performs relatively well for all architectures except for~\mips, for which~\idapro performs exceptionally well.

For CCR and Control Flow Graph analysis (explained next), in order to have a uniform comparison, we selected those functions whose boundaries are correctly identified by all the tools.
Unlike FBD, results are more uniform for CCR (Middle part of~\tbl{tab:tooldifftesting}).
Here, all tools except~\radare have relatively the same number of single tool failures (0.3\% - 3\%). 
These single tool failures reveal interesting issues with these tools. Even the highly rated IDA Pro Decompiler (HexRays) fails to identify the following signature of the function \textbf{"make\_timespec"} in a binary of the \textbf{sleep} program in \textbf{coreutils}.
\\
\\
\textbf{"make\_timespec (time\_t s, long int ns)"}
\\
\\
Whereas all the other tools correctly detect two parameters. 
The large amount of multi-tool failures (\failurem
: 49.70\%) indicates that all the tools fail to accurately detect the calling convention for a large number of functions.
Overall,~\ghidra{} seems to perform relatively well in accurately identifying calling convention (i.e., number of function parameters).

All tools perform equally for calling convention analysis across all architectures. However,
\ghidra performs marginally well compared to other tools.

\begin{figure}[t]
\centering
\includegraphics[width=.5\columnwidth]{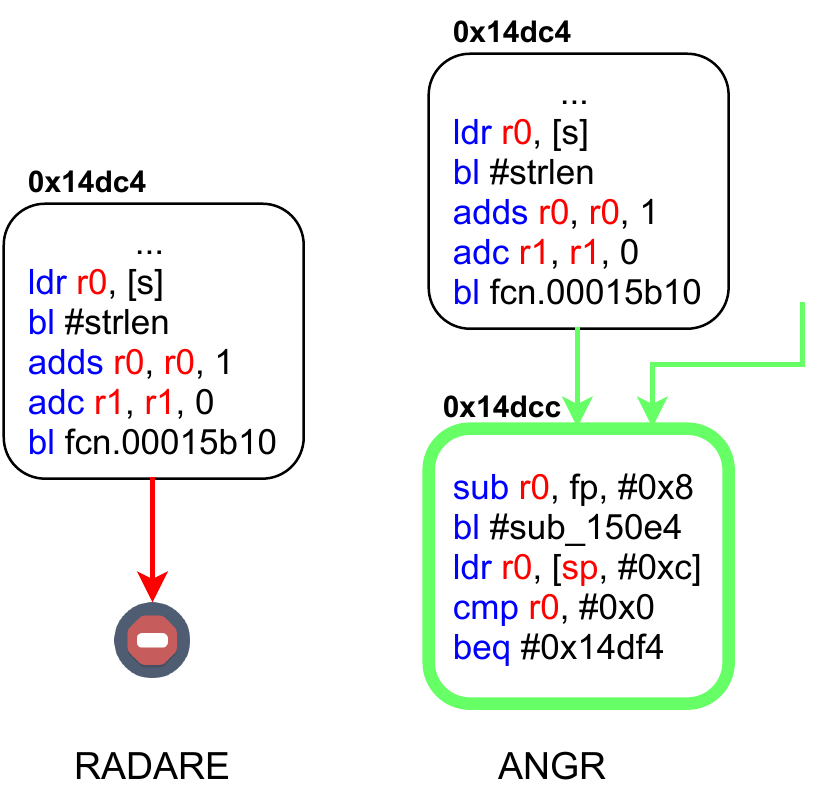}
\caption{CFGs that show the difference in basic blocks between \radare{} and \angr. For \radare{}, there are no basic blocks after the address \textbf{0x14dc4}. Whereas for \angr there are basic blocks after \textbf{0x14dc4}.} 
\label{fig:angr_ida_radare_example}
\end{figure}

\noindent
\textbf{Control Flow Graph (CFG) Recovery: }
This analysis aims to find control flow graphs~\cite{xu2009constructing} of all the functions in a binary.
These graphs contain nodes, commonly called basic blocks, and the edges represent possible control flows in the corresponding function.
Generating ground truth CFG is tricky. Either we need to modify the compiler backend (not generalizable) to emit this information or use one of the binary analysis tools to build it. However, as we presented earlier, these tools might have bugs.
To handle this, we perform differential testing by normalizing the CFG of all the tools to a common format using \textbf{networkx}~\cite{hagberg2020networkx} and comparing them with each other.
The bottom part of~\tbl{tab:tooldifftesting} shows the results.
On average, all tools produce the same or different CFG for 24.28\% and 45.01\% of the functions across all architectures.
Similar to the results of the previous analyses~\radare again performs worse with 20.75\% unique failures.
Although CFG is such a common analysis, it is interesting to see the difference in the results produced by different tools.
We manually inspected a few of these differences and found that most of these are indeed failures.
For instance, for a binary of  \textbf{elfedit} compiled for \arm, \radare{} produced a different result than the rest of the tools.
On further inspection, as shown in~\fig{fig:angr_ida_radare_example}, we find that \radare{} fails to detect the blocks after the address \textbf{0x14dc4} (left side). In comparison,~\angr CFG (right side) accurately detects the blocks after this address.

We have dissected the results further in~\tbl{tab:details_cfg}.
We categorized the divergence of each tool based on the underlying root causes,~\ie Mismatch in the number of basic blocks (\textbf{N}); Number of basic block matches, but the starting addresses differ (\textbf{A}) or the ending address or the size of one or more basic blocks differ (\textbf{S}), or the edges do not match (\textbf{E}); Incomplete output (\ie tool had an internal failure and did not return any basic blocks) (\textbf{P}) and finally timeout (\textbf{T}).

Here we find that most tools diverge on the number of basic blocks for a given function, except for \radare, in which case most divergences are due to incomplete output,~\ie tool failures. 
The \textbf{N} divergences in \idapro for x64 are mostly due to failures in tail-call detection, particularly the reason for almost 50\% of these seem to be stray  \textbf{ud2} instructions after a tail call. Although the number of \textbf{N} divergences looks significant for \angr, further analysis revealed that approximately 99\% of these failures are cases where \angr chooses to merge jumps to the immediate next address with no other edges into the same basic block. Although this deviates from the approach the other tools take, it can be considered a design choice.
Nonetheless, all tools also have internal failures while computing CFG, as indicated by the~\textbf{P} column - these cases represent bugs in the underlying tools and can assist developers in fixing the underlying issues.
We are in the process of organizing these results with appropriate reproducer scripts and reporting them to the corresponding tool developers.


\noindent
\textbf{Summary: } 
The analyzed tools have been previously tested with binaries generated using standard optimization levels~\cite{wang2017ramblr}.
Our results indicate that~\ourtool generates binaries that can effectively reveal issues (missed by regular binaries) in static analysis tools.
Consequently,~\ourtool can be used to supplement the existing binary datasets to test and further improve binary static analysis tools.

\noindent
\textbf{Impact: }
Our results also raise interesting questions about evaluating advanced binary analysis techniques based on the above tools. For instance, Consider OSPREY~\cite{zhang2021osprey}, a recent type inference technique on binaries based on BDA~\cite{zhang2019bda} which uses~\radare for disassembly and CFG. 
Our evaluation shows issues with~\radare for function boundary detection and CFG recovery.
However, OSPREY ignores functions that are missed by~\radare and perform comparative evaluation on~\idapro and~\ghidra and show that OSPREY performs better on those functions detected by all these.
However, the functions detected by all these tools, which, as we show in our evaluation (\sect{subsub:funcbound}) is considerably less.
This raises questions about the actual effectiveness of OSPREY as \idapro and~\ghidra may be better or worse on functions missed by~\radare.
This problem becomes severe when we compare similar techniques built using different binary analysis tools.
We strongly suggest that binary analysis research should pay particular attention to comparative evaluation, especially when using different binary analysis tools.

\noindent
\textbf{Tool Crashes.}
We also found that~\angr and~\idapro~\emph{crashed} on certain binaries.
\emph{Specifically,~\angr crashed on 263 binaries and~\idapro on one binary}.
For~\angr, the crashes are in their python framework, whereas for~\idapro, the crash is in the \textbf{libdwarf} library.
All our issues have been reported and acknowledged by corresponding developers. These issues are being actively fixed. 

%

{
\tiny
\begin{table*}[h]
\centering

\resizebox{\textwidth}{!}{
\begin{tabular}{lr|rrrrrrrrrrrr}
\toprule
\multicolumn{1}{c|}{\multirow{3}{*}{\textbf{Arch.}}} &
  \multicolumn{1}{c|}{\multirow{3}{*}{ \begin{tabular}[c]{@{}c@{}}\textbf{Randomly} \\ \textbf{Sampled} \\ \textbf{Binaries}\end{tabular}}} &
  \multicolumn{12}{c}{ \textbf{Function Boundary Detection}} \\ \cline{3-14}
  
\multicolumn{1}{c|}{} &
  \multicolumn{1}{c|}{} &
  \multicolumn{8}{c|}{\failures} &
  \multicolumn{2}{c|}{\multirow{2}{*}{\failurem}} &
  \multicolumn{2}{c}{\multirow{2}{*}{\successa}} \\ \cline{3-10} 
  
\multicolumn{1}{c|}{} &
  \multicolumn{1}{c|}{} &
  \multicolumn{2}{c}{\angr} &
  \multicolumn{2}{c}{\ghidra} &
  \multicolumn{2}{c}{\idapro} &
  \multicolumn{2}{c|}{\radare} &
  \multicolumn{2}{c|}{} &
  \multicolumn{2}{c}{} \\ \cline{3-12}
   \hline

  \rowcolor{black!15} \multicolumn{1}{l|}{x86} &
  4,600 &
  \multicolumn{2}{r|}{\hlgreen{{271 (5.89\%)}}} &
  \multicolumn{2}{r|}{492 (10.7\%)} &
  \multicolumn{2}{r|}{1,019 (22.15\%)} &
  \multicolumn{2}{r|}{\hlred{4,588 (99.74\%)}} &
  \multicolumn{2}{r|}{1,160 (25.22\%)} &   
  \multicolumn{2}{r}{12 (0.26\%)}  \\

  \multicolumn{1}{l|}{x64} &
  3,516 &
  \multicolumn{2}{r|}{\hlgreen{213 (6.06\%)}} &
  \multicolumn{2}{r|}{302 (8.59\%)} &
  \multicolumn{2}{r|}{255 (7.25\%)} &
  \multicolumn{2}{r|}{\hlred{2,349 (66.81\%)}} &
  \multicolumn{2}{r|}{563 (16.01\%)} & 
  \multicolumn{2}{r}{1,092 (31.06\%)} \\

  \rowcolor{black!15} \multicolumn{1}{l|}{ARM} &
  5,382 &
  \multicolumn{2}{r|}{\hlgreen{2,388 (44.37\%)}} &
  \multicolumn{2}{r|}{2,891 (53.72\%)} &
  \multicolumn{2}{r|}{2,872 (53.36\%)} &
  \multicolumn{2}{r|}{\hlred{3,324 (61.76\%)}} &
  \multicolumn{2}{r|}{3,020 (56.11\%)} & 
  \multicolumn{2}{r}{1,485 (27.59\%)} \\

  \multicolumn{1}{l|}{MIPS} &
  4,818 &
  \multicolumn{2}{r|}{2,744 (56.95\%)} &
  \multicolumn{2}{r|}{2,149 (44.6\%)} &
  \multicolumn{2}{r|}{\hlgreen{679 (14.09\%)}} &
  \multicolumn{2}{r|}{\hlred{3,750 (77.83\%)}} &
  \multicolumn{2}{r|}{2,977 (61.79\%)} & 
  \multicolumn{2}{r}{1,068 (22.17\%)} \\

  \hline
  
  \rowcolor{black!15} \multicolumn{1}{l|}{\textbf{Total}} &
  18,316 &
  \multicolumn{2}{r|}{5,616 (30.66\%)} &
  \multicolumn{2}{r|}{5,834 (31.85\%)} &
  \multicolumn{2}{r|}{\hlgreen{4,825 (26.34\%)}} &
  \multicolumn{2}{r|}{\hlred{14,011 (76.50\%)}} &
  \multicolumn{2}{r|}{7,720 (42.15\%)} & 
  \multicolumn{2}{r}{3,657 (19.97\%)}
   \\  

\bottomrule
\end{tabular}
}

\resizebox{\textwidth}{!}{
\begin{tabular}{lr|rrrrrrrrrrrr}
\toprule
\multicolumn{1}{c|}{\multirow{3}{*}{\textbf{Arch.}}} &
  \multicolumn{1}{c|}{\multirow{3}{*}{ \begin{tabular}[c]{@{}c@{}}\textbf{Total} \\ \textbf{No. of} \\ \textbf{Functions}*\end{tabular}}} &
  \multicolumn{12}{c}{ \textbf{Calling Convention Recovery}} \\ \cline{3-14}
  
\multicolumn{1}{c|}{} &
  \multicolumn{1}{c|}{} &
  \multicolumn{8}{c|}{\failures} &
  \multicolumn{2}{c|}{\multirow{2}{*}{\failurem}} &
  \multicolumn{2}{c}{\multirow{2}{*}{\successa}} \\ \cline{3-10} 
  
\multicolumn{1}{c|}{} &
  \multicolumn{1}{c|}{} &
  \multicolumn{2}{c}{\angr} &
  \multicolumn{2}{c}{\ghidra} &
  \multicolumn{2}{c}{\idapro} &
  \multicolumn{2}{c|}{\radare} &
  \multicolumn{2}{c|}{} &
  \multicolumn{2}{c}{} \\ \cline{3-10}
  
 \hline

  \rowcolor{black!15} \multicolumn{1}{l|}{x86} &
  49,546 &
  \multicolumn{2}{r|}{\hlgreen{85 (0.17\%)}} &
  \multicolumn{2}{r|}{299 (0.60\%)} &
  \multicolumn{2}{r|}{740 (1.49\%)} &
  \multicolumn{2}{r|}{\hlred{9,162 (18.49\%)}} &
  \multicolumn{2}{r|}{18,295 (36.92\%)} & 
  \multicolumn{2}{r}{20,965 (42.31\%)} \\

  \multicolumn{1}{l|}{x64} &
  80,174 &
  \multicolumn{2}{r|}{1,732 (2.16\%)} &
  \multicolumn{2}{r|}{\hlgreen{470 (0.58\%)}} &
  \multicolumn{2}{r|}{4,521 (5.64\%)} &
  \multicolumn{2}{r|}{\hlred{16,018 (19.98\%)}} &
  \multicolumn{2}{r|}{17,695 (22.07\%)} & 
  \multicolumn{2}{r}{39,738 (49.56\%)} \\

  \rowcolor{black!15} \multicolumn{1}{l|}{ARM} &
  228,107 &
  \multicolumn{2}{r|}{13,906 (6.10\%)} &
  \multicolumn{2}{r|}{\hlgreen{816 (0.36\%)}} &
  \multicolumn{2}{r|}{5,012 (2.20\%)} &
  \multicolumn{2}{r|}{\hlred{34,892 (15.30\%)}} &
  \multicolumn{2}{r|}{141,680 (62.11\%)} & 
  \multicolumn{2}{r}{31,801 (13.94\%)} \\

  \multicolumn{1}{l|}{MIPS} &
  132,461 &
  \multicolumn{2}{r|}{390 (0.29\%)} &
  \multicolumn{2}{r|}{\hlgreen{129 (0.10\%)}} &
  \multicolumn{2}{r|}{356 (0.27\%)} &
  \multicolumn{2}{r|}{\hlred{57,729 (43.58\%)}} &
  \multicolumn{2}{r|}{66,028 (49.85\%)} & 
  \multicolumn{2}{r}{7,829 (5.91\%)} \\

  \hline
  
  \rowcolor{black!15} \multicolumn{1}{l|}{\textbf{Total}} &
  490,288 &
  \multicolumn{2}{r|}{16,113 (3.29 \%)} &
  \multicolumn{2}{r|}{\hlgreen{1,714 (0.35\%)}} &
  \multicolumn{2}{r|}{10,629 (2.17\%)} &
  \multicolumn{2}{r|}{\hlred{117,801 (24.03\%)}} &
  \multicolumn{2}{r|}{243,698 (49.70\%)} & 
  \multicolumn{2}{r}{100,333 (20.46\%)} \\  

\bottomrule
\end{tabular}
}

\resizebox{\textwidth}{!}{
\begin{tabular}{lr|rrrrrrrrrrrr}
\toprule
\multicolumn{1}{c|}{\multirow{3}{*}{\textbf{Arch.}}} &
  \multicolumn{1}{c|}{\multirow{3}{*}{ \begin{tabular}[c]{@{}c@{}}\textbf{Total} \\ \textbf{No. of} \\ \textbf{Functions*}\end{tabular}}} &
  \multicolumn{12}{c}{ \textbf{Control Flow Graph Recovery}} \\ \cline{3-14}
  
\multicolumn{1}{c|}{} &
  \multicolumn{1}{c|}{} &
  \multicolumn{8}{c|}{\failures} &
  \multicolumn{2}{c|}{\multirow{2}{*}{\failurem}} &
  \multicolumn{2}{c}{\multirow{2}{*}{\successa}} \\ \cline{3-10} 
  
\multicolumn{1}{c|}{} &
  \multicolumn{1}{c|}{} &
  \multicolumn{2}{c}{\angr} &
  \multicolumn{2}{c}{\ghidra} &
  \multicolumn{2}{c}{\idapro} &
  \multicolumn{2}{c|}{\radare} &
  \multicolumn{2}{c|}{} &
  \multicolumn{2}{c}{} \\ \cline{3-10}
  
 \hline

  \rowcolor{black!15} \multicolumn{1}{l|}{x86} &
  121,108 &
  \multicolumn{2}{r|}{10,622 (8.77\%)} &
  \multicolumn{2}{r|}{\hlgreen{75 (0.06\%)}} &
  \multicolumn{2}{r|}{375 (0.31\%)} &
  \multicolumn{2}{r|}{\hlred{26,209 (21.64\%)}} &
  \multicolumn{2}{r|}{48,146 (39.75\%)} & 
  \multicolumn{2}{r}{35,681 (29.46\%)} \\

  \multicolumn{1}{l|}{x64} &
  109,791 &
  \multicolumn{2}{r|}{5,025 (4.58\%)} &
  \multicolumn{2}{r|}{\hlgreen{153 (0.14\%)}} &
  \multicolumn{2}{r|}{3,108 (2.83\%)} &
  \multicolumn{2}{r|}{\hlred{36,303 (33.07\%)}} &
  \multicolumn{2}{r|}{27,629 (25.17\%)} & 
  \multicolumn{2}{r}{37,573 (34.22\%)} \\

  \rowcolor{black!15} \multicolumn{1}{l|}{ARM} &
  105,674 &
  \multicolumn{2}{r|}{8,182 (7.74\%)} &
  \multicolumn{2}{r|}{91 (0.09\%)} &
  \multicolumn{2}{r|}{\hlgreen{79 (0.07\%)}} &
  \multicolumn{2}{r|}{\hlred{16,414 (15.53\%)}} &
  \multicolumn{2}{r|}{60,534 (57.28\%)} & 
  \multicolumn{2}{r}{20,374 (19.28\%)} \\

  \multicolumn{1}{l|}{MIPS} &
  126,179 &
  \multicolumn{2}{r|}{\hlred{18,244 (14.46\%)}} &
  \multicolumn{2}{r|}{\hlgreen{53 (0.04\%)}} &
  \multicolumn{2}{r|}{64 (0.05\%)} &
  \multicolumn{2}{r|}{17,115 (13.56\%)} &
  \multicolumn{2}{r|}{71,988 (57.05\%)} & 
  \multicolumn{2}{r}{ 18,712 (14.83\%)} \\

  \hline
  
  \rowcolor{black!15} \multicolumn{1}{l|}{\textbf{Total}} &
  462,752 &
  \multicolumn{2}{r|}{42,073 (9.09\%)} &
  \multicolumn{2}{r|}{\hlgreen{372 (0.08\%)}} &
  \multicolumn{2}{r|}{3,626 (0.78\%)} &
  \multicolumn{2}{r|}{\hlred{96,041 (20.75\%)}} &
  \multicolumn{2}{r|}{208,297 (45.01\%)} & 
  \multicolumn{2}{r}{112,340 (24.28\%)} \\ 

\bottomrule
\end{tabular}
}

\caption{Results of differential testing of various analysis. For function boundary detection and calling convention recovery,~\failures and~\failurem indicate the number of binaries with single tool divergence (\ie only one tool produces a different result) and multi-tool divergence (\ie Multiple tools produce different results), respectively. ~\successa shows the number of times all the tools perfectly agreed with each other. For control flow graph recovery,~\failures, ~\failurem indicate the divergence for number of functions ~\successa indicates the number of times all tools agree on functions. (*) Functions in the randomly sampled binaries whose boundaries are correctly identified by all the tools.}
\label{tab:tooldifftesting}
\end{table*}
}

{
\tiny
\begin{table*}[h]
  \centering
  \setlength{\extrarowheight}{0pt}
  \addtolength{\extrarowheight}{\aboverulesep}
  \addtolength{\extrarowheight}{\belowrulesep}
  \setlength{\aboverulesep}{0pt}
  \setlength{\belowrulesep}{0pt}
  \resizebox{\textwidth}{!}{\begin{tabular}{l|r|r|r|r|r|r|r|r|r|r|r|r|r|r|r|r|r|r|r|r|r|r|r|r}
      \toprule
      \multirow{2}{*}{\textbf{Arch.}}        & \multicolumn{6}{c|}{\textbf{\angr}}                              & \multicolumn{6}{c|}{\textbf{\ghidra}}                           & \multicolumn{6}{c|}{\textbf{\idapro}}                              & \multicolumn{6}{c}{\textbf{\radare}}                                 \\
      \cline{2-6}\cmidrule{7-25}
                                         & \textbf{N} & \textbf{A~} & \textbf{S} & \textbf{E} & \textbf{P} & \textbf{T} & \textbf{N} & \textbf{A} & \textbf{S} & \textbf{E} & \textbf{P} & \textbf{T} & \textbf{N} & \textbf{A} & \textbf{S} & \textbf{E} & \textbf{P} & \textbf{T} & \textbf{N} & \textbf{A} & \textbf{S} & \textbf{E} & \textbf{P} & \textbf{T} \\
      \midrule
      \rowcolor{black!15} x86            & 10,389      & 0           & 3          & 10         & 220        & 0           & 18         & 0          & 0          & 0          & 57         &  0          & 304        & 0          & 10         & 28         & 0         &  33          & 5,091       & 24         & 0          & 0          & 21,094      &  0         \\
      x64                                & 3,863       & 0           & 12         & 16         & 1,134       & 0           & 94         & 0          & 7          & 0          & 52         & 0           & 2,844       & 4          & 87         & 173        & 0          & 0           & 6,246       & 84         & 0          & 0          & 29,973      &  0          \\
      \rowcolor{black!15} ARM            & 7,678       & 0           & 0          & 0          & 233        & 271           & 6          & 0          & 0          & 0          & 85         & 0           & 23         & 0          & 11         & 23         & 22         & 0           & 8,920       & 30         & 45         & 0          & 7,419       & 0           \\
      MIPS                               & 18,244      & 0           & 0          & 0          & 0          &  0          & 0          & 0          & 18         & 0          & 35         & 0           & 46         & 0          & 7          & 11         & 0          & 0           & 1,989       & 0          & 0          & 1          & 15,125      &  0         \\
      \midrule
      \rowcolor{black!15} \textbf{Total} & 40,174      & 0           & 15         & 26         & 1,587       &  271          & 118        & 0          & 25         & 0          & 229        & 0           & 3,217       & 4          & 115        & 235        & 22         &  33          & 22,246      & 138        & 45         & 1          & 73,611      &   0         \\
      \bottomrule
    \end{tabular}}
  \caption{Detailed breakdown of the CFG results with single tool divergence. 'N' indicates a mismatch in the number of basic blocks. 'A' shows cases where the number of basic blocks match, but the starting addresses differ. 'S' indicates cases where the size(s) of the basic blocks do not match, 'E' indicates cases where the edges do not match, and 'P' indicates cases when the tools gave incomplete output. 'T' indicates cases for which the tool timed out.}
  \label{tab:details_cfg}
\end{table*}
}

\subsection{Applicability to Test~\ac{ML} tools}
\label{subsub:robusttest}
In this section, we will explore the second application of testing the robustness of~\ac{ML} tools on the binaries generated by~\ourtool{}.
We selected the following two recent tools, as these are open-source and claim to have high accuracy.

\noindent
\textbf{Binary diffing techniques (\asmvec~\cite{ding2019asm2vec} and \safe~\cite{massarelli2019safe}): }
These are representation learning techniques based on neural networks.
They propose a representation of binaries into a vector space such that binaries will be close.
In other words, the distance between the vector representations of two should be minimal, ideally 0.
As in these papers, we use cosine similarity to measure the difference between the generated vectors. Specifically, we compute ~\emph{Inverse cosine similarity} (\ie 1 - cosine similarity) denoted as~\cosinesim; a large value of~\cosinesim indicates a higher difference.
Ideally, we would want the~\cosinesim to be very low for all the binaries for the same program. However, this is not the case.
We got the pre-trained models for these two tools and used the corresponding vectors to compute the \cosinesim of the generated binaries.
Our results as shown that~\emph{\ourtool{} was able to generate binaries with higher~\cosinesim scores than O3 for all the programs}. 
A detailed analysis of results is shown in
\ifreport
~\apdx{apdx:bindifftools}
\fi
.
This shows that~\ourtool{} can generate binaries that cannot be detected as similar by the existing techniques.
We suggest that these techniques should use~\ourtool{} to improve their dataset, which could help in building more accurate models.

\noindent
\textbf{Debug information prediction (\debin~\cite{he2018debin}): }
This technique combines two complementary probabilistic models to predict~\emph{types of variables in a stripped binary}.
Their evaluation shows that on average,~\debin has an F1 score of 67\%.
We used their pre-trained model and tested its accuracy on each of the binaries generated by~\ourtool{}.
The~\tbl{tab:debintable} shows the results of our experiment.
Although ~\debin uses binaries of different optimization levels to train their model, it still performs extremely poorly on the binaries generated by~\ourtool{}, with F1 score dropping to 12.9\% (x86), 18.2\% (x64) and, 13.6\% (ARM) from the reported 67\%. 
We tried to use \stateformer~\cite{pei2021stateformer}, a recent learning-based tool to predict types. However, the pre-built model and the dataset are inaccessible and did not receive help from the authors as well. 
Nonetheless, our results on other ML techniques show that existing approaches to generate binary datasets are inadequate and~\ourtool{} can help to improve existing datasets, consequently helping in creating better models.

\begin{table}[h]
\centering
\footnotesize
\begin{tabular}{c||c||rr||rr||rr}
\toprule
\multicolumn{1}{c||}{\multirow{2}{*}{\textbf{Arch}}} & \multicolumn{1}{c||}{} & \multicolumn{2}{c||}{\textbf{Prec}} & \multicolumn{2}{c||}{\textbf{Rec}} & \multicolumn{2}{c}{\textbf{F1}} \\ \cline{2-8}
\multicolumn{1}{c||}{}                                       & \multicolumn{1}{c||}{} & \multicolumn{1}{c|}{\textbf{R}} & \multicolumn{1}{c||}{\textbf{O}} & \multicolumn{1}{c|}{\textbf{R}} & \multicolumn{1}{c||}{\textbf{O}} & \multicolumn{1}{c|}{\textbf{R}} & \multicolumn{1}{c}{\textbf{O}}                                                                                      \\ \midrule

\rowcolor{black!15}\textbf{\xeightsix}   & Name    &  62.6      &   \hlred{7.4}    &       62.5        &  \hlred{15.6}        &   62.5    &  \hlred{10.0}    \\
                   \textbf{}             & Type    &  63.7      &  \hlred{11.1}    &       63.7        &  \hlred{33.9}        &   63.7    &  \hlred{15.6}    \\
                   \textbf{}             & Overall &  63.1      &   \hlred{9.3}    &       63.1        &  \hlred{24.0}        &   63.1    &  \hlred{12.9}    \\

\rowcolor{black!15}\textbf{\xsixfour}    & Name    &  63.5      &  \hlred{3.2}     &       63.1        &   \hlred{5.2}        &   63.3    &   \hlred{3.9}    \\
                   \textbf{}             & Type    &  74.1      & \hlred{24.7}     &       73.4        &  \hlred{47.8}        &   73.8    &  \hlred{31.9}    \\
                   \textbf{}             & Overall &  68.8      & \hlred{14.2}     &       68.3        &  \hlred{26.7}        &   68.6    &  \hlred{18.2}    \\
                   
\rowcolor{black!15}\textbf{\arm}         & Name    &  61.6      &  \hlred{7.0}     &       61.3        &  \hlred{12.5}        &   61.5    &  \hlred{8.7}     \\
                   \textbf{}             & Type    &  66.8      & \hlred{14.6}     &       68.0        &  \hlred{24.8}        &   67.4    &  \hlred{17.9}    \\
                   \textbf{}             & Overall &  64.2      & \hlred{10.7}     &       64.7        &  \hlred{20.3}        &   64.5    &  \hlred{13.6}    \\

\bottomrule

\end{tabular}
\caption{Figure showing~\textbf{Precision (Prec)},~\textbf{Recall (Rec)}, and,~\textbf{F1} scores for~\debin across 3 different architectures: The columns~\textbf{Reported (R)} and~\textbf{Observed (O)} show reported and observed scores on~\ourtool generated binaries.}
\label{tab:debintable}
\end{table}
\section{Limitations and Future work}
\label{sec:limitation}
Although,~\ourtool{} is effective at generating a plethora of binaries. It has the following limitations.\\
\noindent
\textbf{Compiler bugs:} We assume that the provided compiler preserves the semantics of the program in the generated binary. However, this may not be the case. 
The compiler may have bugs~\cite{yang2011finding, sun2016toward} resulting in binaries that may not be semantically equivalent, especially those concerning undefined behavior. \\
\noindent
\textbf{Compiler frontend overhead:} Although we mainly use the backend or code generation component of a compiler, in the general case, we unnecessarily run all components of the compiler, including its frontend.
This adds a lot of overhead~\cite{li2010lightweight} as demonstrated by the relatively low yield by~\gccversion (\sect{subsub:bingeneffectiveness}).

\noindent
\textbf{Completeness of the Generated Dataset:} \ourtool{} uses existing programs to generate diverse binaries, and the completeness (\eg instructions covered in the underlying~\ac{ISA}) of the generated dataset depends on the features present in the corresponding programs. For instance, a program that does not use any floating point variables is unlikely to produce binaries with floating point instructions~\eg~\textbf{fcmovb}.
This can be handled by using programs from diverse sources, such as Debian Repositories~\cite{galindo2010debian}, GitHub, etc. We can also use systematic approaches such as Csmith~\cite{yang2011finding} to generate C programs with the desired features and then use them in~\ourtool{} to generate a complete binary dataset.

\noindent
\textbf{Minimizing compiler crashes:} Although, as discussed in~\sect{subsubsec:compilercrashes},~\ourtool could find compiler crashes, it does not try to triage (\eg minimizing options and the target binary) them.
We plan to integrate techniques like Delta Debugging~\cite{zeller2002simplifying} in our future work to minimize the set of crash-causing compiler flags.

\vspace{-2pt}
\section{Related Work}
\label{sec:relatedwork}
Program obfuscation~\cite{nagra2009surreptitious} is a well-known technique to change a program's structure without affecting the underlying functionality.
One possible approach to the problem of this paper is to use various
obfuscation techniques~\cite{wang2017turing, lan2017lambda} to generate semantically equivalent 
but structurally different binaries.
Many initial techniques~\cite{ceccato2009effectiveness, viticchie2016assessment} are aimed towards source or IR level obfuscation.
\tigress~\cite{tigress} is a source-to-source transformer that has various configurable transformations, such as control-flow flattening~\cite{laszlo2009obfuscating} and opaque-predicates~\cite{collberg1998manufacturing}.
Similarly, \obfusllvm~\cite{junod2015obfuscator} enables applying a limited set of transformations 
at the LLVM IR level.
Closure~\cite{liu2017stochastic} uses stochastic optimization to select a sequence of transformations to produce the optimal obfuscation potency.
Although these techniques are effective at modifying the program at IR or source code level,
they have less impact on the generated binary~\cite{madou2006effectiveness}.

A few binary-level techniques obfuscate control-flow using error handling semantics such as signals~\cite{popov2007binary} and exception handling~\cite{10.1145/1866307.1866368}.
Other virtual machine--based techniques~\cite{fang2011multi, kochberger2021sok} transform the given binary into a custom virtual machine.
These binary-level techniques are known to introduce performance overhead~\cite{anckaert2007program}.
The binary-level techniques are based on a fixed set of carefully designed patterns~\cite{nagra2009surreptitious}, which do not capture the entire range of behaviors of the underlying~\ac{ISA}.
Finally, the primary goal of obfuscation techniques is to generate a hard-to-understand version of a given program~\cite{barak2001possibility}. 
In contrast,~\ourtool{} does not care about the understandability of the generated binary as 
long as it is different from all previously seen variations.
The use of a compiler to generate binaries have been explored before, especially in the area of software diversity~\cite{schaefer2012software, larsen2014sok, jackson2011compiler, HOSSEINZADEH201872}.
These techniques only consider limited, non-performance-impacting transformations.
\ourtool{} has no such restrictions and explores all possible variations of the
binary using compiler flags.

Although the effects of non-standard compiler optimizations on the generated binary have been explored before~\cite{blackmore2017automatically}, the recent work~\bintuner{}~\cite{ren2021unleashing} is the most closely related to~\ourtool{}.
However, as explained in~\sect{sec:introduction},~\bintuner{} requires considerable effort to use as it requires specifying conflicting compiler flags manually as first-order constraints.
~\ourtool{} is completely automated and uses a feedback-guided approach to identify conflicting options automatically.
Furthermore, as shown in~\sect{subsub:cornucopavsbintuner},~\ourtool{} is more effective than~\bintuner{} in efficiently generating diverse binaries.
%
The use of fuzzing, especially~\AFL{}, to generate a sequence of tokens has been explored before to fuzz interpreters~\cite{salls2021token}. 
Our approach allows the fuzzer to use its input generation ability fully, and
enables~\ourtool{} to be easily configurable to use other fuzzers.

\section{Conclusions}
\label{sec:conclusion}
We present~\ourtool{}, an architecture, compiler agnostic and automated framework that generates a plethora of diverse binaries from program source code by using feedback-guided fuzzing. Our evaluation shows that~\ourtool{} is generally more effective at generating diverse binaries for a given program than~\bintuner, a closely related work. It can be scaled on multiple threads for faster binary generation and better resource utilization. We showed that many binary analysis frameworks perform poorly on ~\ourtool{} generated binaries opening up opportunities for more research in this area. We envision that~\ourtool{} becomes part of a binary analysis testing framework and helps in creating more robust analysis tools.

\section{Acknowledgments}
\label{sec:conclusion}
We would like to thank all the anonymous reviewers for their valuable feedback and suggestions which helped in making this paper the best version of itself. This project was supported in part by the Defense Advanced Research Projects Agency (DARPA) under contract number N6600120C4031 and National Science Foundation (NSF) awards CCF-1908504 and CCF-1919197. The U.S. Government is authorized to reproduce and distribute reprints for Governmental purposes notwithstanding any copyright notation thereon. Any opinions, findings, conclusions or recommendations expressed in this material are those of the author(s) and do not necessarily reflect the views of United States Government, National Science Foundation or any agency thereof.

\newpage
\Urlmuskip=0mu plus 1mu
\bibliographystyle{ACM-Reference-Format}
\bibliography{references}

\ifreport
\appendix
\section{Appendix}

\section{Size of programs v/s number of binaries generated:}
\label{appendix:sizevsnumber}
We also want to evaluate how the program's size affects the
number of binaries generated.
We consider the size of~\textbf{O0} bitcode file as the program size.

Intuitively, large programs should have more optimization opportunities, and hence~\ourtool{} should produce more binaries.
However, large programs have relatively higher compilation times.
As we fix the fuzzing time (\fuzztime hours), the large programs have fewer compilations than smaller programs.
\fig{fig:sizevsvarients} shows the~\ac{PDF} of number of binaries (y-axis) generated for programs of corresponding size (x-axis).
For multiple programs of the same size, we consider the average number of binaries across these programs. As we can see, the number of binaries increases with the program size showing the effect of increased optimization opportunities for smaller sizes. However, the number of binaries decreases as the size increases for larger sizes (> 18-21 KB) with increased compilation time, which reduces the number of compilations resulting in a smaller number of binaries.

\begin{figure}[h]

\begin{tikzpicture}

\begin{axis}[
	xlabel=Size (KB),
	ylabel=Number of Variants,
	ymin=0,
	ylabel near ticks,
	xlabel near ticks,
	ymax=4000,
	xmode=log,
	xtick={7, 14, 21, 28, 35, 42, 49, 85, 135, 175, 225},
	xticklabels={7, 14, 21, 28, 35, 42, 49, 85, 135, 175, 225},
	ytick={500,1000,...,4000},
	xmin=0,
	xmax=225,
	width=9cm,height=6cm,
    yticklabel style={font=\small},
	xticklabel style={font=\small, rotate=70},
	grid=both,
    legend style={at={(0.72,0.95)},anchor=north, font=\footnotesize},
    legend columns=3,
    ]
    

\addplot[color=top1,mark=x] table[col sep=comma] {data/variationsizepdf/llvm/data_x86.txt};\addlegendentry{x86};
\addplot[color=bot3,mark=square] table[col sep=comma] {data/variationsizepdf/llvm/data_x64.txt};\addlegendentry{x64};
\addplot[color=bot2,mark=triangle] table[col sep=comma] {data/variationsizepdf/llvm/data_arm.txt};\addlegendentry{arm};
\addplot[color=bot1,mark=diamond] table[col sep=comma] {data/variationsizepdf/llvm/data_mips.txt};\addlegendentry{mips};



\end{axis}
\end{tikzpicture}

\caption{PDF showing percentage of binaries generated for programs of different sizes across various architecture and compiler combinations.}
\label{fig:sizevsvarients}
\end{figure}
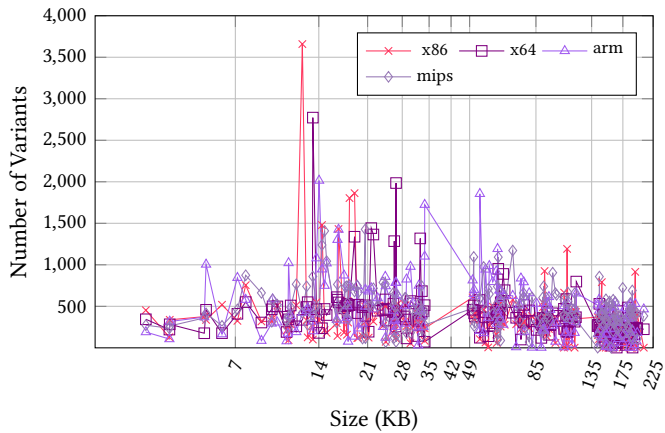

\section{Quality of the generated binaries (Detailed analysis)}
\label{sec:qualgenbin}
In addition to many binaries, we also want~\ourtool to generate different varieties of binaries.
We use~\textit{Inverse Bindiff} (\bindiff) scores (\ie 1 - similarity (bindiff) score) to measure the variance or quality of the generated binaries. 
A large~\bindiff score indicates a higher difference.
The~\fig{fig:bindiffllvmcdf} shows the CDF the percentage of programs against~\bindiff scores.
Each line indicates the CDF of~\bindiff scores in comparison with the binary generated by the corresponding optimization option.
For instance, the Avg-O0 line in all subfigures of \fig{fig:bindiffllvmcdf} shows the trend of average~\bindiff scores of all binaries generated for a program against the binary compiled with O0.
Specifically, a point (x, y) on a Avg-O0 line shows the average~\bindiff score (of binaries generated by~\ourtool in comparison with O0) for y\% of the programs is less than or equal to x.
Similarly, Med-O0, Max-O0 show the trend for median and maximum~\bindiff scores.
Furthermore, the *-O3 shows the comparison with O3 (maximum level) compiled binary.
We also show the trend of \bindiff scores of O0 against other optimization levels in O0 v/s Ox lines.

\noindent
\textbf{Maximum \bindiff scores:}
The Max-* lines, which are consistently at the right of O0 v/s O3, indicates that~\emph{~\ourtool{} was able to generate binaries with higher~\bindiff scores than O3 for all the programs}.
For instance, for x64 (\fig{fig:bindiffxsixfour}), the maximum~\bindiff score of O0 v/s O3 is 0.4 as indicated by the peak of the line. However, there are at least 43\% (100-57) of programs for which the maximum~\bindiff score of~\ourtool{} generated binaries ranges from 0.45 - 0.7.
(Inferred from two points (0.45, 57) and (0.7, 100) on the Max - O0 line).
Furthermore, as shown by the Max-O3 lines, the binaries generated by~\ourtool have a consistently higher score than O3 binaries hinting that we could explore the optimization options that are not covered in O3.

\noindent
\textbf{Average \bindiff scores:}
The average~\bindiff scores of~\ourtool{} generated binaries for x86 (\fig{fig:bindiffxeightsix}), and x64 (\fig{fig:bindiffxsixfour}) is slightly lower than that of Avg-O0 v/s O0 v/s Ox variants. 
This is expected and good as~\ourtool{} is able to explore the binary variations that are not explored by existing optimization levels.
It is interesting to see that the average scores are more in the case of ARM (\fig{fig:bindiffarm}) and MIPS (\fig{fig:bindiffmips}). This is also because of the underlying~\ac{RISC} ISA and the greater optimization it provides.
Finally, as expected, the median \bindiff scores also follow the same trend as average.

This shows that~\ourtool{} is effective at generating a large number of different binaries and can explore the variants that are not covered by the standard optimization levels~\ie O0, O1, O2, and, O3.

\definecolor{avgcolor}{rgb}{0.59, 0.44, 0.84}
\definecolor{mediancolor}{rgb}{0.34, 0.01, 0.1}
\definecolor{maxcolor}{rgb}{0.0, 0.0, 0.61}

\definecolor{O0vsO1}{rgb}{0.55, 0.71, 0.0}
\definecolor{O0vsO2}{rgb}{1.0, 0.77, 0.05}
\definecolor{O0vsO3}{rgb}{0.89, 0.35, 0.13}

\begin{figure*}[h]
\begin{minipage}{0.25\textwidth}
\centering
 \resizebox{0.7\textwidth}{!}{
\begin{tikzpicture}
\begin{axis}[
	xlabel=\bindiff score,
	ylabel=Percentage of Programs,
	ytick style={font=\footnotesize},
	xtick style={font=\footnotesize},
	ymin=0,
	ylabel near ticks,
	xlabel near ticks,
	ymax=100,
	xtick={0, 0.05, 0.1, 0.15, 0.2, 0.25, 0.3, 0.35, 0.4, 0.45, 0.5, 0.55, 0.6, 0.65, 0.7, 0.75, 0.8, 0.85, 0.9, 0.95, 1.0},
	ytick={10,20,...,100},
    yticklabel style={font=\footnotesize},
	xticklabel style={font=\footnotesize, rotate=60},
	xmin=0,
	xmax=1.0,
	grid=both,
    legend style={at={(0.5,-0.22)},anchor=north, font=\tiny},
    legend columns=3,
    ]

\addplot[color=avgcolor,mark=square*] table[col sep=comma] {data/bindiff/llvm/x86/average.txt};\addlegendentry{Avg-O0};
\addplot[color=mediancolor,mark=triangle*] table[col sep=comma] {data/bindiff/llvm/x86/median.txt};\addlegendentry{Med-O0};
\addplot[color=maxcolor,mark=*] table[col sep=comma] {data/bindiff/llvm/x86/max.txt};\addlegendentry{Max-O0};

\addplot[color=avgcolor,mark=square] table[col sep=comma] {data/bindiff/llvm/x86/average03.txt};\addlegendentry{Avg-O3};
\addplot[color=mediancolor,mark=triangle] table[col sep=comma] {data/bindiff/llvm/x86/median03.txt};\addlegendentry{Med-O3};
\addplot[color=maxcolor,mark=+] table[col sep=comma] {data/bindiff/llvm/x86/max03.txt};\addlegendentry{Max-O3};

\addplot[color=O0vsO1,mark=square] table[col sep=comma] {data/bindiff/llvm/x86/O0vsO1.txt};\addlegendentry{O0 vs O1};
\addplot[color=O0vsO2,mark=triangle] table[col sep=comma] {data/bindiff/llvm/x86/O0vsO2.txt};\addlegendentry{O0 vs O2};
\addplot[color=O0vsO3,mark=+] table[col sep=comma] {data/bindiff/llvm/x86/O0vsO3.txt};\addlegendentry{O0 vs O3};

\end{axis}
\end{tikzpicture}
  }
\caption{\xeightsix}
\label{fig:bindiffxeightsix}
\end{minipage}%
\begin{minipage}{0.25\textwidth}
 \centering
   \resizebox{0.7\textwidth}{!}{
    \begin{tikzpicture}
\begin{axis}[
	xlabel=\bindiff score,
	yticklabels={,,},
	xtick style={font=\footnotesize},
	ymin=0,
	ylabel near ticks,
	xlabel near ticks,
	ymax=100,
	xtick={0, 0.05, 0.1, 0.15, 0.2, 0.25, 0.3, 0.35, 0.4, 0.45, 0.5, 0.55, 0.6, 0.65, 0.7, 0.75, 0.8, 0.85, 0.9, 0.95, 1.0},
	ytick={10,20,...,100},
    yticklabel style={font=\footnotesize},
	xticklabel style={font=\footnotesize, rotate=60},
	xmin=0,
	xmax=1.0,
	grid=both,
    legend style={at={(0.5,-0.22)},anchor=north, font=\tiny},
    legend columns=3,
    ]

\addplot[color=avgcolor,mark=square*] table[col sep=comma] {data/bindiff/llvm/x64/average.txt};\addlegendentry{Avg-O0};
\addplot[color=mediancolor,mark=triangle*] table[col sep=comma] {data/bindiff/llvm/x64/median.txt};\addlegendentry{Med-O0};
\addplot[color=maxcolor,mark=*] table[col sep=comma] {data/bindiff/llvm/x64/max.txt};\addlegendentry{Max-O0};

\addplot[color=avgcolor,mark=square] table[col sep=comma] {data/bindiff/llvm/x64/average03.txt};\addlegendentry{Avg-O3};
\addplot[color=mediancolor,mark=triangle] table[col sep=comma] {data/bindiff/llvm/x64/median03.txt};\addlegendentry{Med-O3};
\addplot[color=maxcolor,mark=+] table[col sep=comma] {data/bindiff/llvm/x64/max03.txt};\addlegendentry{Max-O3};

\addplot[color=O0vsO1,mark=square] table[col sep=comma] {data/bindiff/llvm/x64/O0vsO1.txt};\addlegendentry{O0 vs O1};
\addplot[color=O0vsO2,mark=triangle] table[col sep=comma] {data/bindiff/llvm/x64/O0vsO2.txt};\addlegendentry{O0 vs O2};
\addplot[color=O0vsO3,mark=+] table[col sep=comma] {data/bindiff/llvm/x64/O0vsO3.txt};\addlegendentry{O0 vs O3};

\end{axis}
\end{tikzpicture}
    }
    \caption{\xsixfour}
    \label{fig:bindiffxsixfour}

\end{minipage}%
\begin{minipage}{0.25\textwidth}
  \resizebox{0.7\textwidth}{!}{
    \begin{tikzpicture}
\begin{axis}[
	xlabel=\bindiff score,
	yticklabels={,,},
	xtick style={font=\footnotesize},
	ymin=0,
	ylabel near ticks,
	xlabel near ticks,
	ymax=100,
	xtick={0, 0.05, 0.1, 0.15, 0.2, 0.25, 0.3, 0.35, 0.4, 0.45, 0.5, 0.55, 0.6, 0.65, 0.7, 0.75, 0.8, 0.85, 0.9, 0.95, 1.0},
	ytick={10,20,...,100},
    yticklabel style={font=\footnotesize},
	xticklabel style={font=\footnotesize, rotate=60},
	xmin=0,
	xmax=1.0,
	grid=both,
    legend style={at={(0.5,-0.22)},anchor=north, font=\tiny},
    legend columns=3,
    ]

\addplot[color=avgcolor,mark=square*] table[col sep=comma] {data/bindiff/llvm/arm/average.txt};\addlegendentry{Avg-O0};
\addplot[color=mediancolor,mark=triangle*] table[col sep=comma] {data/bindiff/llvm/arm/median.txt};\addlegendentry{Med-O0};
\addplot[color=maxcolor,mark=*] table[col sep=comma] {data/bindiff/llvm/arm/max.txt};\addlegendentry{Max-O0};

\addplot[color=avgcolor,mark=square] table[col sep=comma] {data/bindiff/llvm/arm/average03.txt};\addlegendentry{Avg-O3};
\addplot[color=mediancolor,mark=triangle] table[col sep=comma] {data/bindiff/llvm/arm/median03.txt};\addlegendentry{Med-O3};
\addplot[color=maxcolor,mark=+] table[col sep=comma] {data/bindiff/llvm/arm/max03.txt};\addlegendentry{Max-O3};

\addplot[color=O0vsO1,mark=square] table[col sep=comma] {data/bindiff/llvm/arm/O0vsO1.txt};\addlegendentry{O0 vs O1};
\addplot[color=O0vsO2,mark=triangle] table[col sep=comma] {data/bindiff/llvm/arm/O0vsO2.txt};\addlegendentry{O0 vs O2};
\addplot[color=O0vsO3,mark=+] table[col sep=comma] {data/bindiff/llvm/arm/O0vsO3.txt};\addlegendentry{O0 vs O3};

\end{axis}
\end{tikzpicture}
     }
    \caption{\arm}
    \label{fig:bindiffarm}
\end{minipage}%
\begin{minipage}{0.25\textwidth}
     \resizebox{0.7\textwidth}{!}{
    \begin{tikzpicture}
\begin{axis}[
	xlabel=\bindiff score,
	yticklabels={,,},
	xtick style={font=\footnotesize},
	ymin=0,
	ylabel near ticks,
	xlabel near ticks,
	ymax=100,
	xtick={0, 0.05, 0.1, 0.15, 0.2, 0.25, 0.3, 0.35, 0.4, 0.45, 0.5, 0.55, 0.6, 0.65, 0.7, 0.75, 0.8, 0.85, 0.9, 0.95, 1.0},
	ytick={10,20,...,100},
    yticklabel style={font=\footnotesize},
	xticklabel style={font=\footnotesize, rotate=60},
	xmin=0,
	xmax=1.0,
	grid=both,
    legend style={at={(0.5,-0.22)},anchor=north, font=\tiny},
    legend columns=3,
    ]

\addplot[color=avgcolor,mark=square*] table[col sep=comma] {data/bindiff/llvm/mips/average.txt};\addlegendentry{Avg-O0};
\addplot[color=mediancolor,mark=triangle*] table[col sep=comma] {data/bindiff/llvm/mips/median.txt};\addlegendentry{Med-O0};
\addplot[color=maxcolor,mark=*] table[col sep=comma] {data/bindiff/llvm/mips/max.txt};\addlegendentry{Max-O0};

\addplot[color=avgcolor,mark=square] table[col sep=comma] {data/bindiff/llvm/mips/average03.txt};\addlegendentry{Avg-O3};
\addplot[color=mediancolor,mark=triangle] table[col sep=comma] {data/bindiff/llvm/mips/median03.txt};\addlegendentry{Med-O3};
\addplot[color=maxcolor,mark=+] table[col sep=comma] {data/bindiff/llvm/mips/max03.txt};\addlegendentry{Max-O3};

\addplot[color=O0vsO1,mark=square] table[col sep=comma] {data/bindiff/llvm/mips/O0vsO1.txt};\addlegendentry{O0 vs O1};
\addplot[color=O0vsO2,mark=triangle] table[col sep=comma] {data/bindiff/llvm/mips/O0vsO2.txt};\addlegendentry{O0 vs O2};
\addplot[color=O0vsO3,mark=+] table[col sep=comma] {data/bindiff/llvm/mips/O0vsO3.txt};\addlegendentry{O0 vs O3};

\end{axis}
\end{tikzpicture}
     }
    \caption{\mips}
    \label{fig:bindiffmips}
\end{minipage}
\caption{CDF of~\textit{Inverse Bindiff} (\bindiff) scores of generated binaries for all programs across different architectures.}
\label{fig:bindiffllvmcdf}

\end{figure*}

\begin{figure*}[h]

        \centering

    \begin{minipage}{0.25\textwidth}
     \centering

      \resizebox{0.7\textwidth}{!}{
    \begin{tikzpicture}
        \begin{axis}[
            xlabel=Number of Binary Variants,
            ylabel=Percentage of Programs,
            ylabel near ticks,
            xlabel near ticks,
            symbolic x coords={0-100,100-200,200-300,300-400,400-500,500-600,600-700,700-3700},
            width=8cm,height=6cm,
            yticklabel style={font=\footnotesize},
	        xticklabel style={font=\footnotesize, rotate=45},
            ymin=0,
            ymax=30,
            ymajorgrids=true,
            xtick=data
          ]
          \addplot[ybar,draw=thircolor, pattern=horizontal lines] table[col sep=comma] {data/variations/llvm/data_x32.txt};
        \end{axis}
\end{tikzpicture}
   }
    
\caption{\xeightsix}
\label{subfig:varclangeightsix}

\end{minipage}%
\begin{minipage}{0.25\textwidth}
 \centering
   \resizebox{0.7\textwidth}{!}{
    \begin{tikzpicture}
        \begin{axis}[
            xlabel=Number of Binary Variants,
            yticklabels={,,},
            xlabel near ticks,
            symbolic x coords={0-100,100-200,200-300,300-400,400-500,500-600,600-700,700-2800},
            width=8cm,height=6cm,
            yticklabel style={font=\footnotesize},
	        xticklabel style={font=\footnotesize, rotate=45},
            ymin=0,
            ymax=30,
            ymajorgrids=true,
            xtick=data
          ]
          \addplot[ybar,draw=sixcolor, pattern=vertical lines] table[col sep=comma] {data/variations/llvm/data_x64.txt};
        \end{axis}
\end{tikzpicture}
       }
        
    \caption{\xsixfour}
    \label{subfig:varclangsixfour}

\end{minipage}%
\begin{minipage}{0.25\textwidth}
  \resizebox{0.7\textwidth}{!}{
    \begin{tikzpicture}
        \begin{axis}[
            xlabel=Number of Binary Variants,
            yticklabels={,,},
            xlabel near ticks,
            symbolic x coords={0-100,100-200,200-300,300-400,400-500,500-600,600-700,700-800,800-900,900-1000,1000-1100,1100-2100},
            width=8cm,height=6cm,
	        xticklabel style={font=\footnotesize, rotate=45},
            ymin=0,
            ymax=30,
            ymajorgrids=true,
            xtick=data
          ]
          \addplot[ybar,draw=armcolor, pattern=grid] table[col sep=comma] {data/variations/llvm/data_arm.txt};
        \end{axis}
\end{tikzpicture}
     }
        
    \caption{\arm}
    \label{subfig:varclangarm}

\end{minipage}%
\begin{minipage}{0.25\textwidth}
  \resizebox{0.7\textwidth}{!}{
    \begin{tikzpicture}
        \begin{axis}[
            xlabel=Number of Binary Variants,
            yticklabels={,,},
            xlabel near ticks,
            symbolic x coords={0-100,100-200,200-300,300-400,400-500,500-600,600-700,700-800,800-900,900-1500},
            width=8cm,height=6cm,
            yticklabel style={font=\footnotesize},
	        xticklabel style={font=\footnotesize, rotate=45},
            ymin=0,
            ymax=30,
            ymajorgrids=true,
            xtick=data
          ]
          \addplot[ybar, draw=mipscolor, pattern=dots] table[col sep=comma] {data/variations/llvm/data_mips.txt};
        \end{axis}
\end{tikzpicture}
      }
        
    \caption{\mips}
    \label{subfig:varclangmips}

\end{minipage}

\caption{Number of binaries generated vs percentage of programs that lie in that particular range for the different architectures.}
\label{fig:varclangtotal}

\end{figure*}

\begin{figure}
\texttt{
 \\ *** Scheduling failed! ***
 \\ ...
 \\ t54: ch,glue = CopyToReg t0, 
 \\ Register:i32 \$eflags, t46:1, rapper.c:481:31
 \\ has not been scheduled!
 \\ ...
 \\ has not been scheduled!
 \\ SU(6): PHYS REG COPY
 \\ has successors left!
}
\caption{One of the crashes found in optimization scheduler of LLVM}
\label{lst:llvmcompcrash}
\end{figure}

\section{Extensibility}
\label{apdx:extensibility}
Our implementation of~\ourtool{} is extensible in the following three aspects. 
\begin{itemize}
\item\textbf{Adding a new compiler.} Users can configure~\ourtool{} to use a custom compiler by just changing the compiler environment
variable of the binary generator to the file system path of the compiler and providing the list of all possible compiler flags in a file.
\item\textbf{Adding a custom fitness checking function.} As mentioned before, users interested in developing
custom fitness checking functions can modify our existing python based \fitserver or need to expose an
interface that accepts a binary and returns a~\ac{DScore} (a floating-point number).
\item\textbf{Adding a new source package.} We support all packages that can be built using \textbf{configure} and~\textbf{make} (or~\textbf{cmake}). Users just need to provide the path to the package (\ie\textbf{.tar.gz}).
\end{itemize}

\section{Testing ML Based Binary Diffing Techniques}
\label{apdx:bindifftools}
\asmvec supports only x64, whereas as~\safe supports both x64 and ARM.
These tools claim that their model can detect semantically equivalence, such that binaries compiled with different optimization levels (\ie Ox) will be very similar or have a low~\cosinesim score.
We got the pre-trained models for these two tools and used the corresponding vectors to compute the \cosinesim of the generated binaries.
The~\fig{fig:asm2vecgcc} shows the CDF of the average (Avg-*), median (Med-*), and maximum (Max-*)~\cosinesim score for all the binaries generated for each program against the binary generated with the corresponding optimization (Similar to~\sect{subsub:bingeneffectiveness}). 

Both the tools claim that the comparison of O0 with O3 is toughest, i.e., should have the highest score. 
However, as shown in~\fig{fig:asm2vecgcc} by the Max-* lines, which are consistently at the right of O0 v/s O3, indicates that~\emph{\ourtool{} was able to generate binaries with higher~\cosinesim scores than O3 for all the programs}. 
The trend is the same for average and median as well across all architectures for both~\asmvec and~\safe.
This shows that~\ourtool{} can generate binaries that cannot be detected as similar by the existing techniques.
We suggest that these techniques should use~\ourtool{} to improve their dataset, which could help in creating better models.
\definecolor{avgcolor}{rgb}{0.59, 0.44, 0.84}
\definecolor{mediancolor}{rgb}{0.34, 0.01, 0.1}
\definecolor{maxcolor}{rgb}{0.0, 0.0, 0.61}

\definecolor{O0vsO1}{rgb}{0.55, 0.71, 0.0}
\definecolor{O0vsO2}{rgb}{1.0, 0.77, 0.05}
\definecolor{O0vsO3}{rgb}{0.89, 0.35, 0.13}

\begin{figure*}[h]
\begin{minipage}{0.33\textwidth}
\centering
\resizebox{0.7\textwidth}{!}{
\begin{tikzpicture}
\begin{axis}[
	xlabel=\asmvec score,
	ylabel=Percentage of Programs,
	ytick style={font=\footnotesize},
	xtick style={font=\footnotesize},
	ymin=0,
	ylabel near ticks,
	xlabel near ticks,
	ymax=100,
	xtick={0, 0.05, 0.1, 0.15, 0.2, 0.25, 0.3, 0.35, 0.4, 0.45, 0.5, 0.55, 0.6, 0.65, 0.7, 0.75, 0.8, 0.85, 0.9, 0.95, 1.0, 1.05, 1.10, 1.15, 1.2},
	ytick={10,20,...,100},
    yticklabel style={font=\footnotesize},
	xticklabel style={font=\footnotesize, rotate=60},
	xmin=0,
	xmax=1.2,
	width=9cm,height=6cm,
	grid=both,
    legend style={at={(0.5,-0.22)},anchor=north, font=\tiny},
    legend columns=3,
    ]

\addplot[color=avgcolor,mark=square*] table[col sep=comma] {data/asm2vec/llvm/x64/average.txt};\addlegendentry{Avg-O0};
\addplot[color=mediancolor,mark=triangle*] table[col sep=comma] {data/asm2vec/llvm/x64/median.txt};\addlegendentry{Med-O0};
\addplot[color=maxcolor,mark=*] table[col sep=comma] {data/asm2vec/llvm/x64/max.txt};\addlegendentry{Max-O0};

\addplot[color=avgcolor,mark=square] table[col sep=comma] {data/asm2vec/llvm/x64/averageO3.txt};\addlegendentry{Avg-O3};
\addplot[color=mediancolor,mark=triangle] table[col sep=comma] {data/asm2vec/llvm/x64/medianO3.txt};\addlegendentry{Med-O3};
\addplot[color=maxcolor,mark=+] table[col sep=comma] {data/asm2vec/llvm/x64/maxO3.txt};\addlegendentry{Max-O3};
\addplot[color=O0vsO3,mark=+] table[col sep=comma] {data/asm2vec/llvm/x64/O0vsO3.txt};\addlegendentry{O0 vs O3};

\end{axis}
\end{tikzpicture}
}
\caption{CDF of \asmvec score}
\end{minipage}
\begin{minipage}{0.33\textwidth}
\centering
  \resizebox{0.7\textwidth}{!}{
\begin{tikzpicture}

\begin{axis}[
	xlabel=\safe score,
	ylabel=Percentage of Programs,
	ytick style={font=\footnotesize},
	xtick style={font=\footnotesize},
	ymin=0,
	ylabel near ticks,
	xlabel near ticks,
	ymax=100,
	xtick={0, 0.05, 0.1, 0.15, 0.2, 0.25, 0.3, 0.35, 0.4, 0.45, 0.5, 0.55, 0.6, 0.65, 0.7, 0.75, 0.8, 0.85, 0.9, 0.95, 1.0},
	ytick={10,20,...,100},
    yticklabel style={font=\footnotesize},
	xticklabel style={font=\footnotesize, rotate=60},
	xmin=0,
	xmax=1,
	width=9cm,height=6cm,
	grid=both,
    legend style={at={(0.5,-0.22)},anchor=north, font=\tiny},
    legend columns=3,
    ]

\addplot[color=avgcolor,mark=square*] table[col sep=comma] {data/safe/llvm/x64/average.txt};\addlegendentry{Avg-O0};
\addplot[color=mediancolor,mark=triangle*] table[col sep=comma] {data/safe/llvm/x64/median.txt};\addlegendentry{Med-O0};
\addplot[color=maxcolor,mark=*] table[col sep=comma] {data/safe/llvm/x64/max.txt};\addlegendentry{Max-O0};

\addplot[color=avgcolor,mark=square] table[col sep=comma] {data/safe/llvm/x64/averageO3.txt};\addlegendentry{Avg-O3};
\addplot[color=mediancolor,mark=triangle] table[col sep=comma] {data/safe/llvm/x64/medianO3.txt};\addlegendentry{Med-O3};
\addplot[color=maxcolor,mark=+] table[col sep=comma] {data/safe/llvm/x64/maxO3.txt};\addlegendentry{Max-O3};
\addplot[color=O0vsO3,mark=+] table[col sep=comma] {data/safe/llvm/x64/O0vsO3.txt};\addlegendentry{O0 vs O3};
\end{axis}
\end{tikzpicture}
}
\caption{CDF of \safe score}
\end{minipage}
\begin{minipage}{0.33\textwidth}
\centering
  \resizebox{0.7\textwidth}{!}{
\begin{tikzpicture}

\begin{axis}[
	xlabel=\safe ARM score,
	xlabel style={font=\footnotesize},
	ylabel=Percentage of Programs,
	ylabel style={font=\footnotesize},
	ytick style={font=\footnotesize},
	xtick style={font=\footnotesize},
	ymin=0,
	ylabel near ticks,
	xlabel near ticks,
	ymax=100,
	xtick={0, 0.05, 0.1, 0.15, 0.2, 0.25, 0.3, 0.35, 0.4, 0.45, 0.5, 0.55, 0.6, 0.65, 0.7, 0.75, 0.8, 0.85, 0.9, 0.95, 1.0},
	ytick={10,20,...,100},
    yticklabel style={font=\footnotesize},
	xticklabel style={font=\footnotesize, rotate=60},
	xmin=0,
	xmax=1,
	width=9cm,height=6cm,
	grid=both,
    legend style={at={(0.5,-0.22)},anchor=north, font=\tiny},
    legend columns=3,
    ]

We display the CDF of safe cosine similarity scores.
\addplot[color=avgcolor,mark=square*] table[col sep=comma] {data/safe/llvm/arm/average.txt};\addlegendentry{Avg-O0};
\addplot[color=mediancolor,mark=triangle*] table[col sep=comma] {data/safe/llvm/arm/median.txt};\addlegendentry{Med-O0};
\addplot[color=maxcolor,mark=*] table[col sep=comma] {data/safe/llvm/arm/max.txt};\addlegendentry{Max-O0};

\addplot[color=avgcolor,mark=square] table[col sep=comma] {data/safe/llvm/arm/averageO3.txt};\addlegendentry{Avg-O3};
\addplot[color=mediancolor,mark=triangle] table[col sep=comma] {data/safe/llvm/arm/medianO3.txt};\addlegendentry{Med-O3};
\addplot[color=maxcolor,mark=+] table[col sep=comma] {data/safe/llvm/arm/maxO3.txt};\addlegendentry{Max-O3};
\addplot[color=O0vsO3,mark=+] table[col sep=comma] {data/safe/llvm/arm/O0vsO3.txt};\addlegendentry{O0 vs O3};
\end{axis}
\end{tikzpicture}
}
\caption{CDF for ARM binaries (\safe).}
\end{minipage}
\caption{CDFs of scores generated by different tools on binaries generated by llvm for all programs for~\xsixfour and~\arm.}
\label{fig:asm2vecgcc}
\end{figure*}

\section{\ourtool in parallel mode}
\label{apdx:parellmodecornu}

\begin{figure*}
\includegraphics[scale=0.35]{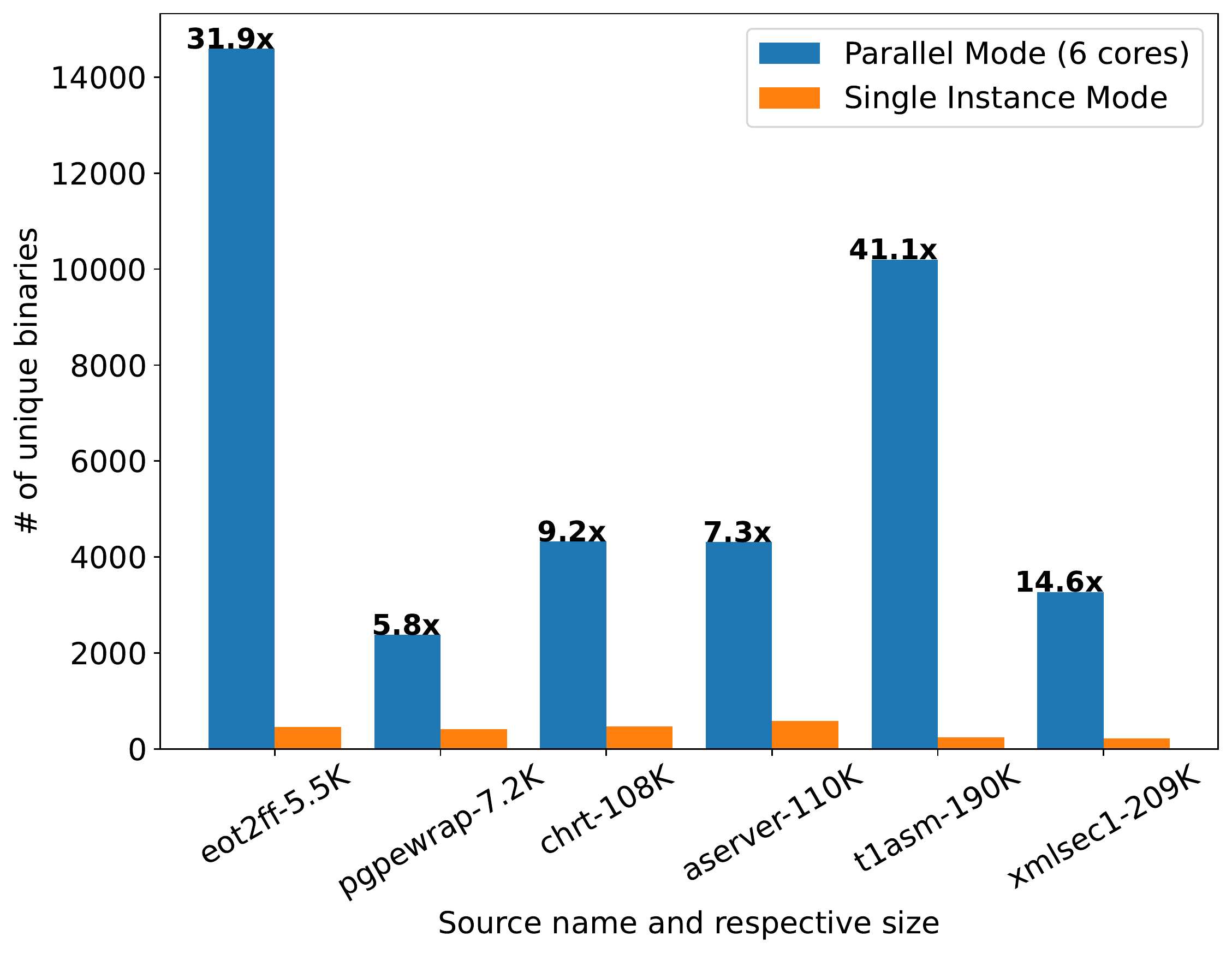}

\caption{Number of binaries generated for a particular source file when fuzzed in multi-core mode vs in single core mode, each source was fuzzed for 6 hours in both modes}
\label{fig:afl-parallel}
\end{figure*}

\section{Extending to Different Compilers}
\label{apdx:differentcompilers}
To demonstrate extensibility of~\ourtool{} in accommodating different compilers, we tested~\ourtool with \gccversion{} version 9.3.0.
As mentioned before, for~\clangversion{}, we optimize our binary generator by avoiding running the frontend and running our optimizations directly on the bitcode files.
But, in the case of~\gccversion{} (with no middle-end optimizer like \clangversion) we run~\ourtool{} as is,~\ie the entire source package will be re-compiled using the selected flags.
Furthermore, each source package can generate multiple binaries. To handle this, we used the number of cores proportional to the number of binaries. For example, for a package with eight binaries, our binary generator compiles it parallelly (\ie{}~\textbf{-j n}) on eight cores.
Similar to \clangversion{}, we executed the \gccversion{} configured \ourtool{} on~\xsixfour for six hours each on several debian source packages containing total of 307 programs.
We got 1,554 binaries, a relatively fewer number of binaries compared to~\clangversion.
The primary reason for this is~\emph{less number of iterations}.
As mentioned before,~\gccversion works directly on the source package, and each fuzzing run takes a relatively long time as we need to first run~\textbf{./configure} and then build.
However,~\clangversion works directly on the bitcode file and takes relatively less time for a run.
Since we used the same fuzzing time (six hours) for both,~\clangversion performs more iterations ($\sim$ 10x more) and has a greater opportunity to produce different binaries.
In fact, when we increased the fuzzing time for~\gccversion, we noticed an increase in the number of generated binaries.
These experiments show that~\ourtool is extensible and can be used with other compilers too.

\fi

\end{document}